\documentclass[12pt, a4paper]{article}
\usepackage{epsf}
\usepackage{cite}
\usepackage{amsmath,amssymb}
\input{colordvi.tex}
\usepackage[usenames,dvipsnames]{color}
\usepackage{graphicx}
\usepackage{ascmac}
\usepackage{hyperref}
\usepackage{slashed}
\usepackage{tikz-feynman}
\counterwithin{equation}{section}
\tikzfeynmanset{compat=1.0.0}

\begin{document}

\begin{titlepage}

\begin{center}
\hfill TU-1304\\
\hfill KEK-QUP-2026-0007
\vskip 1.in

\renewcommand{\thefootnote}{\fnsymbol{footnote}}

{\Large \bf
Anisotropy of Cosmic Background Photons from \\ \vspace{2.5mm}
Annihilating/Decaying Dark Matter
}

\vskip .5in

{\large
Ryosuke Kasuya$^{(a)}$\footnote{kasuya.ryosuke.s5@tohoku.ac.jp}
and
Kazunori Nakayama$^{(a,b)}$\footnote{kazunori.nakayama.d3@tohoku.ac.jp}
}

\vskip 0.5in

$^{(a)}${\em 
Department of Physics, Tohoku University, Sendai 980-8578, Japan
}

\vskip 0.2in

$^{(b)}${\em 
International Center for Quantum-field Measurement Systems for Studies of the Universe and Particles (QUP), KEK, 1-1 Oho, Tsukuba, Ibaraki 305-0801, Japan
}

\end{center}
\vskip .5in

\begin{abstract}

We provide a detailed formulation for calculating the angular power spectrum of the cosmic background photons arising from the dark matter decay or annihilation in a comprehensive manner.
We pay particular attention to the case of dark matter decaying or annihilating into line photons. It is pointed out that taking account of the energy resolution of the detector is essential to correctly evaluate the angular power spectrum.
We apply our formulation to the observational data from infrared, optical, X-ray and gamma-ray telescopes.

\end{abstract}

\end{titlepage}

\tableofcontents

\renewcommand{\thefootnote}{\arabic{footnote}}
\setcounter{footnote}{0}

\section{Introduction}

Searching for dark matter through various cosmic rays, including photons, electrons, positrons, anti-protons and neutrinos, has been one of the promising ways to discover dark matter signals (see, e.g., Ref.~\cite{Klasen:2015uma} for a review).

Among many dark matter models proposed so far, we focus on dark matter decaying or annihilating into photons.
Well-motivated decaying dark matter candidates include a scalar particle like axion, axion-like particle, dilaton/moduli, which decays into two photons, and a fermionic particle like sterile neutrino, which decays into photon plus active neutrino, or others~\cite{Cirelli:2024ssz}.
These dark matter candidates are not absolutely stable and have finite lifetime, but thanks to the smallness of their mass and/or couplings to other particles the lifetime can be much longer than the age of the universe. Only a small fraction of dark matter decays until the present universe, but its signals are still detectable.
On the other hand, dark matter often possesses a (discrete) charge that ensures the stability of dark matter particle. 
Even in such a case, dark matter can pair annihilate into the Standard Model particles. Weakly-Interacting-Massive-Particle is a traditional candidate of annihilating dark matter.
If dark matter is lighter than the electron, the dominant annihilation channel may be two photons (as well as neutrinos).

There are several ways to search for dark matter signals through photons: observing photons produced by dark matter in the Milky Way galaxy, dwarf galaxies or from cosmologically distributed one.
The most studies for the cosmological dark matter decay or annihilation so far considered an isotropic component~\cite{Kawasaki:1997ah,Asaka:1998ju,Bergstrom:2001jj,Ullio:2002pj,Kawasaki:2009nr}, including directly produced photons from dark matter as well as those produced by inverse-Compton scattering off the background photons by high energy electrons and positrons~\cite{Ishiwata:2009dk,Profumo:2009uf}.
However, dark matter is not uniformly distributed in the universe and the dark matter density fluctuation leads to anisotropic signatures for decay/annihilation-produced photons.
Such studies have been done in Refs.~\cite{Ando:2005xg,Ando:2006cr,Ando:2013ff,Fornengo:2013rga,Fornasa:2016ohl} for relatively heavy annihilating dark matter that produces continuum photons with energy above $\sim 1$\,GeV, and in Refs.~\cite{Zandanel:2015xca,Gong:2015hke,Kohri:2017oqn,Kalashev:2018bra,Nakayama:2022jza,Carenza:2023qxh} for relatively light dark matter decaying into line photons.
Interestingly, it is found in Refs.~\cite{Kohri:2017oqn,Kalashev:2018bra} that, when dark matter decays into line photons, a naive calculation of the angular power spectrum leads to a divergence in the limit of infinite detector energy resolution and hence it is essential to take into account the finite energy resolution of the detector.
Constraints on axion-like particles for the mass around $\sim 10\,{\rm eV}$ have been derived in Refs.~\cite{Nakayama:2022jza,Carenza:2023qxh} by using the Hubble Space Telescope data.

The purpose of this paper is twofold.
First, we provide a detailed formalism for calculating the angular power spectrum from both annihilating/decaying dark matter. 
Since the case of dark matter annihilation involves extra issues related to dark matter clump, it will be helpful to formulate both the decay and annihilation case in a parallel way.
Also no one has formulated the case of dark matter annihilating into line photons, which must be completed by taking into account the effect of finite energy resolution in a nontrivial way.
Second, we make use of as many observation data of the angular power spectrum of background photons as possible, from infrared/optical to X-ray range.

This paper is organized as follows.
In Sec.~\ref{sec:formalism} we provide a detailed formulation for calculating the angular power spectrum of the photons from dark matter decay or annihilation.
Some basic ingredients required for these calculations are also summarized in Appendix.
In Sec.~\ref{sec:app} we apply our formalism to some observational data and derive constraints on dark matter decay rate and annihilation cross section.
Sec.~\ref{sec:conc} is devoted to conclusion and discussion.

\section{Formalism}
\label{sec:formalism}

Throughout this paper we consider the flat Friedmann-Robertson-Walker universe, in which the line element is given by
\begin{align}
	ds^2 = -dt^2 + a^2(t) d \vec x^2,
\end{align}
where $a(t)$ is the cosmic scale factor. We often express the scale factor in terms of the redshift $z$ as $a(z) = (1+z)^{-1}$. The density parameter of the dark energy (or cosmological constant) and total matter in the present universe are denoted by $\Omega_{\Lambda 0}$, $\Omega_{m0}$, respectively. They satisfy $\Omega_{\Lambda 0} + \Omega_{m0}=1$.

\subsection{Decaying dark matter}

\subsubsection{Isotropic flux}

Let us denote the dark matter mass by $m$ and the decay rate into arbitrary final state $F$ by $\Gamma_F$, respectively.  The dark matter energy density is denoted by $\rho_{\rm DM}(z)$ at the redshift $z$. Its value in the present universe is expressed by $\rho_{\rm DM,0}$.
The mean photon flux averaged over the observation bandwidth is given by
\begin{align}
	\overline I(\omega) = \int d\omega' \left(\omega'\right)^q \mathcal P(\omega') \int_0^\infty dr W(r,\omega''),
	\label{Imean_dec}
\end{align}
where $q=1$ $(2)$ corresponds to the photon number (energy) flux, $\omega''$ is the photon energy at the production, $\omega'$ is the photon energy at the detector so that they satisfy $\omega''=(1+z)\omega'$, $\mathcal P(\omega')$ is the filter function representing the detector response peaked around $\omega'\simeq \omega$, normalized such that
\begin{align}
	\int d\omega' \mathcal P(\omega') = 1,
\end{align}
and\footnote{
	We are interested in the case that the dark matter lifetime is much longer than the present age of the universe ($t_0$). Thus we take the factor $e^{-\Gamma t_0}$ to be unity.
    We also neglect the attenuation of the photon when traveling the cosmological distance~\cite{Chen:2003gz,Slatyer:2009yq}.
    It is justified for the photon energy range that we will be interested in.}
\begin{align}
	W(r,\omega'')= \sum_F\frac{1}{4\pi} \frac{\rho_{\rm DM}(z)}{m (1+z)^3}\, \Gamma_{F}\, \frac{d\mathcal N^{(F)}_\gamma}{d\omega''},
\end{align}
and $r(z)$ is the comoving distance to the redshift $z$, given by
\begin{align}
	r(z) = \int_0^z \frac{dz'}{H(z')}. \label{rz}
\end{align}
Here the Hubble parameter is given by $H(z) = H_0\sqrt{\Omega_{\Lambda 0} + \Omega_{m0}(1+z)^3}$ with $H_0$ being the present Hubble parameter, which is also expressed in terms of the dimensionless quantity $h$ as $H_0= 100\,h\,{\rm km/s/Mpc}$.
In this paper we only consider the universe after the matter-radiation equality and hence the radiation contribution is negligible.
The photon spectrum produced by the dark matter decay into the final state $F$ is represented by $d\mathcal N^{(F)}_\gamma/d\omega''$. In this paper we are interested in the case of line gamma, i.e., $F=2\gamma$ or $F=\gamma+X$ with $X$ being some light particle that is just regarded as a missing energy. Then we can write
\begin{align}
	\frac{d\mathcal N_\gamma}{d\omega''} = \frac{N_\gamma}{2}\delta\left(\omega'' - \frac{m}{2}\right),
	\label{dNdomega}
\end{align}
where $N_\gamma$ is the number of photon produced per decay.
A typical example of $N_\gamma=2$ is axion-like particles~\cite{OHare:2024nmr}, while $N_\gamma=1$ is sterile neutrino~\cite{Boyarsky:2018tvu}. 
In Refs.~\cite{Kohri:2017oqn,Kalashev:2018bra}, an axion-like particle decaying into photon plus dark photon has been considered, which is another example of $N_\gamma=1$.
By substituting (\ref{dNdomega}), we obtain
\begin{align}
	\overline I(\omega) = \int d\omega' \left(\omega'\right)^{q-1} \mathcal P(\omega') \frac{1}{8\pi}\frac{\rho_{\rm DM,0}}{m}\left.\frac{N_\gamma\Gamma}{H(z)}\right|_{1+z=\frac{m}{2\omega'}}.
\end{align}
For $q=1$, the mean flux $\overline I$ has a unit of, e.g., ${\rm [cm^{-2}s^{-1}sr^{-1}]}$, while for $q=2$ it has a unit of, e.g.,  ${\rm [eV\,cm^{-2}s^{-1}sr^{-1}]}$.\footnote{
    Here we list some conversion formula for the unit often used in optical/infrared or radio astronomy literature: $1\,{\rm nW} = 6.24\times10^9\,{\rm eV/s}$, $1\,{\rm Jy}=10^{-26}\,{\rm W/m^2/Hz}=9.48\times 10^7\,{\rm m^{-2}s^{-1}}$. Thus we have $1\,{\rm nW/m^2} =65.8\,{\rm eV\,Jy}$.} 

\begin{figure}\centering
\includegraphics[width=.5\textwidth]{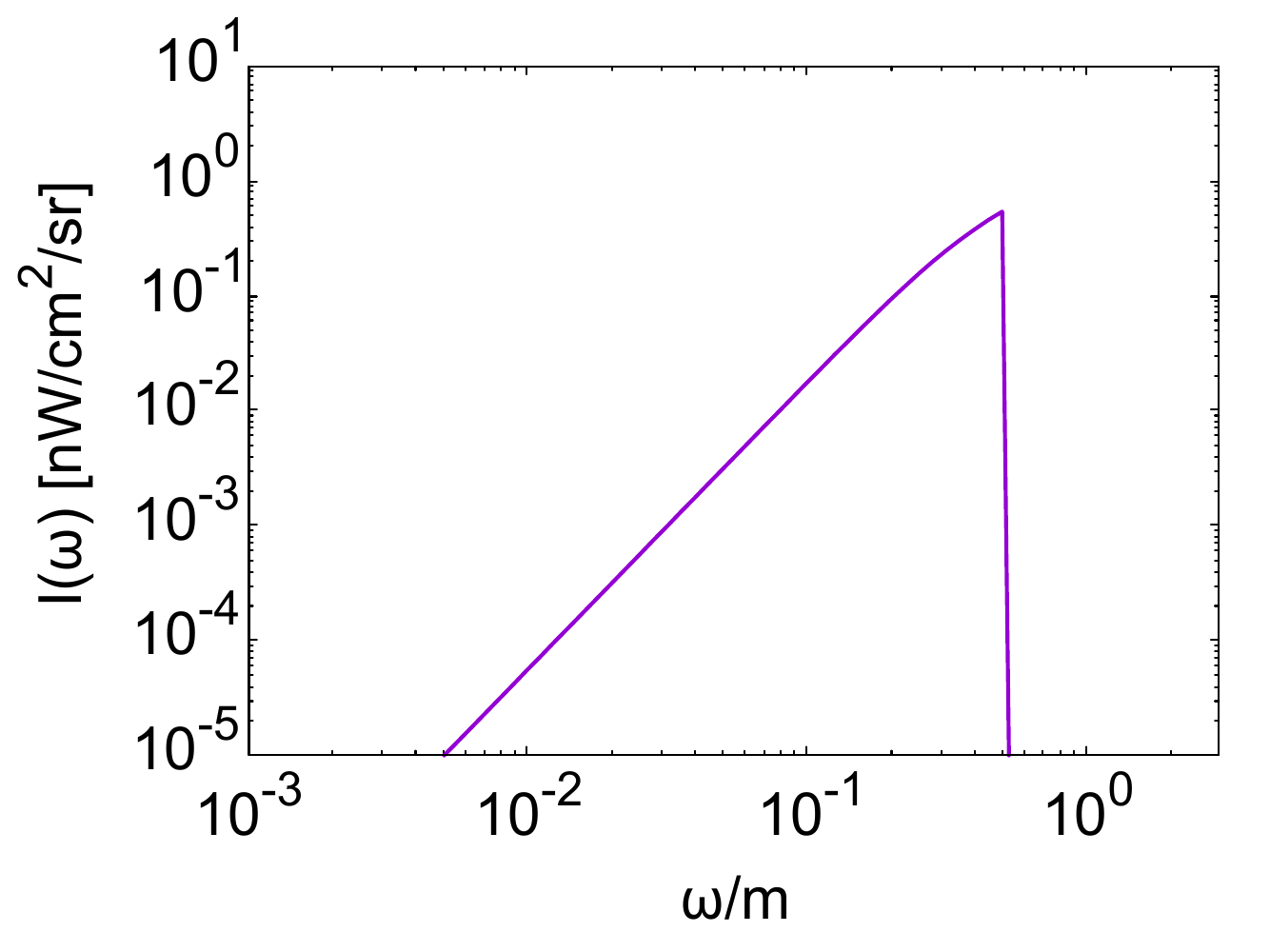}
\caption{The mean intensity of cosmic background photon from dark matter decaying into line photons. We have taken $N_\gamma\Gamma=10^{-20}\,{\rm sec^{-1}}$.}
\label{fig:mean_dec}
\end{figure}

Typically, a detector has a sensitivity at a finite energy range $\omega \pm \Delta\omega/2$ and the sensitivity rapidly falls outside of this range.
A simple approximation is the box shape one:
\begin{align}
	\mathcal P(\omega') = \frac{1}{\Delta\omega}\,\theta\left(\omega'-\omega + \frac{\Delta\omega}{2}\right)\theta\left(\omega + \frac{\Delta\omega}{2}- \omega'\right),
	\label{filter_box}
\end{align}
where $\theta(x)$ denotes the step function. In the narrow band limit we may take $\mathcal P(\omega') = \delta(\omega-\omega')$. In this case we obtain
\begin{align}
	\overline I(\omega) &= \frac{\omega^{q-1}}{8\pi}\frac{\rho_{\rm DM,0}}{m}\left.\frac{N_\gamma\Gamma}{H(z)}\right|_{1+z=\frac{m}{2\omega}} \nonumber\\
    &\simeq 1.1\,[{\rm nW/m^2/sr}] \left(\frac{\omega}{m}\right)\left(\frac{N_\gamma\Gamma}{10^{-24}\,{\rm s^{-1}}}\right)\left[\Omega_{\Lambda 0} + \Omega_{m0}\left(\frac{m}{2\omega}\right)^3\right]^{-1/2}.
\end{align}
In the second line we assumed $q=2$.
Fig.~\ref{fig:mean_dec} shows the mean intensity for 
As we will see in the next subsection, we cannot use the approximation $\mathcal P(\omega') = \delta(\omega-\omega')$ when evaluating the angular power spectrum. 

Note that dark matter actually has nonzero velocity dispersion, hence the line spectrum (\ref{dNdomega}) should be broadened due to the Doppler effect. As far as the detector resolution is worse than the Doppler broadening, which is actually the case in real detectors, the use of  (\ref{dNdomega}) is justified. 
However, in any case, after integration over $z'$ and $\omega'$ in (\ref{Imean_dec}), both the effect of Doppler broadening and detector resolution almost disappear. Thus the mean intensity is not very sensitive to these subtleties. This is not the case for the analysis of angular power spectrum~\cite{Kohri:2017oqn,Kalashev:2018bra}, as we shall see in the next subsection.

\subsubsection{Angular power spectrum}

Taking account of the density fluctuation of dark matter, the actual flux should be anisotropic, i.e., depend on the arrival direction $\hat\Omega = (\theta,\phi)$.
The fluctuation of the flux is given by
\begin{align}
	\delta I(\omega,\hat\Omega) = I (\omega,\hat\Omega) - \overline I(\omega).
\end{align}
We define the normalized (dimensionless) fluctuation as
\begin{align}
	\delta_I(\omega,\hat\Omega) \equiv \frac{\delta I(\omega,\hat\Omega)}{\overline I(\omega)} = \frac{I (\omega,\hat\Omega) - \overline I(\omega)}{\overline I(\omega)}.
\end{align}
Following the standard procedure, it is expanded in terms of the spherical harmonics $Y_{\ell m}(\hat\Omega)$ as
\begin{align}
	\delta_I(\omega,\hat\Omega)  = \sum_{\ell,m} a_{\ell m}(\omega) Y_{\ell m}(\hat\Omega)
	~~~\leftrightarrow~~~ a_{\ell m}(\omega)&=\int d\hat\Omega\,Y^*_{\ell m}(\hat\Omega) \delta_I(\omega,\hat\Omega),
\end{align}
Then the (dimensionless) angular power spectrum is given by
\begin{align}
	C_\ell(\omega) = \left<\left|a_{\ell m}(\omega)\right|^2 \right> = \frac{1}{2\ell+1}\sum_{m=-\ell}^{\ell}\left|a_{\ell m}(\omega)\right|^2.
\end{align}

In order to obtain a concrete expression for $C_\ell$ in terms of the dark matter power spectrum, first note that
\begin{align}
	\delta_I(\omega,\hat\Omega) &= \frac{1}{\overline I} \int d\omega' (\omega')^{q}\mathcal P(\omega') \int dr W(r,\omega'') \delta_\rho(r,\hat\Omega)\\
	&= \frac{1}{\overline I} \int d\omega' (\omega')^{q}\mathcal P(\omega') \int drW(r,\omega'') \int \frac{k^2 dk d\hat\Omega_k}{2\pi^2}\sum_{\ell,m}i^\ell j_\ell(kr)Y_{\ell m}^*(\hat\Omega_k) Y_{\ell m}(\hat\Omega) \delta_\rho(\vec k, z). \nonumber
\end{align}
where we have defined the dark matter density fluctuation as $\delta_\rho\equiv\delta\rho_{\rm DM}/\rho_{\rm DM}$ and used the Rayleigh's formula:
\begin{align}
	e^{i\vec k\cdot \vec x} =4\pi \sum_{\ell,m}i^\ell j_\ell(kr) Y^*_{\ell m}(\hat k) Y_{\ell m}(\hat x).
\end{align}
The angular power spectrum is expressed as
\begin{align}
	C_\ell(\omega) &=\int d\hat\Omega_1 \int d\hat\Omega_2 \, Y^*_{\ell m}(\hat\Omega_1) Y_{\ell m}(\hat\Omega_2)
	\left<\delta_I(\omega,\hat\Omega_1) \delta^*_I(\omega,\hat\Omega_2)\right> \nonumber\\
	&=\frac{1}{\overline I^2(\omega)}\int d\omega_1'(\omega_1')^{q}\mathcal P(\omega_1') \int  d\omega_2' (\omega_2')^{q}\mathcal P(\omega_2') \int dr_1 W(r_1,\omega_1'') \int dr_2 W(r_2, \omega_2'') \nonumber \\
	&\times\int \frac{k_1^2dk_1d\hat\Omega_{k_1}}{2\pi^2}\int \frac{k_2^2dk_2d\hat\Omega_{k_2}}{2\pi^2} j_\ell(k_1r_1) j_\ell(k_2r_2)Y^*_{\ell m}(\hat\Omega_{k_1})Y_{\ell m}(\hat\Omega_{k_2})\left<\delta_\rho(\vec k_1,z_1)\delta^*_\rho(\vec k_2,z_2)\right>  \nonumber\\
	&=\frac{1}{\overline I^2(\omega)}\int d\omega_1' (\omega_1')^{q} \mathcal P(\omega_1') \int d\omega_2' (\omega_2')^{q} \mathcal P(\omega_2')\int \frac{dr_1}{r_1^2}
    W(r_1,\omega_1'')W(r_1,\omega_2'') P_\rho\left(k=\frac{\ell}{r_1}; z_1\right),
	\label{Cl_general}
\end{align}
where $\omega_1''=(1+z)\omega_1'$, $\omega_2''=(1+z)\omega_2'$ and we have used the definition of the matter power spectrum\footnote{
    In this paper, perturbation of the quantity $A$ is represented by $\delta_A(\vec x)$. It should not be confused with the Dirac delta function $\delta(\vec x)$, which does not have any subscript. }
\begin{align}
	\left<\delta_\rho(\vec k_1,z_1)\delta^*_\rho(\vec k_2,z_2)\right>  = (2\pi)^3 \delta(\vec k_1-\vec k_2) P_\rho(k_1; z_1,z_2),
\end{align}
and the Limber approximation~\cite{LoVerde:2008re}
\begin{align}
	\int dk k^2 j_\ell (kr_1) j_\ell(kr_2) P_\rho(k; r_1,r_2) \simeq \frac{\pi}{2} \frac{1}{r_1^2}P_\rho\left(k=\frac{\ell}{r_1}; r_1\right)\delta(r_1-r_2).
    \label{Limber}
\end{align}
For the matter power spectrum $P_\rho(k; z)$, we use the nonlinear one taking account of the halo formation. See Appendix~\ref{app:mat} for detail.

Eq.~(\ref{Cl_general}) is a general expression applicable to any incident photon spectrum $d\mathcal N_\gamma / d\omega''$. 
From now on let us consider the case of line spectrum (\ref{dNdomega}).
One immediately sees that the angular power spectrum (\ref{Cl_general}) diverges in the limit of narrow-band observation $\mathcal P(\omega') \to \delta(\omega-\omega')$~\cite{Kohri:2017oqn}, since the integrand includes four delta functions but there are only three integrals, leaving one delta function.
Therefore, it is essential to take into account the finite detector resolution $\mathcal P(\omega')$~\cite{Kalashev:2018bra,Nakayama:2022jza,Carenza:2023qxh}.
In this case we obtain
\begin{align}
	C_\ell(\omega) = \frac{1}{\overline I^2(\omega)}\int dz \left(\frac{(\omega')^q \mathcal P(\omega')}{1+z}\frac{\rho_{\rm DM,0} \Gamma N_\gamma}{8\pi m} \right)^2 \frac{1}{H(z) r^2(z)} P_\rho\left(k=\frac{\ell}{r(z)}; z\right),
\end{align}
where $\omega' = \frac{m}{2(1+z)}$. This is further rewritten in the form of
\begin{align}
	C_\ell(\omega) =\frac{\int dz \left(\frac{(\omega')^q \mathcal P(\omega')}{(1+z)H(z)}\right)^2 \frac{H(z)}{r^2(z)} P_\rho\left(k=\frac{\ell}{r(z)}; z\right)}{\left[\int dz \frac{(\omega')^q \mathcal P(\omega')}{(1+z)H(z)}  \right]^2}.
    \label{Cl_dec}
\end{align}
This quantity is independent of the dark matter parameters and hence we can make model-independent predictions on $C_\ell$.
The only required information is observation frequency $\omega$ and its bandwidth characterized by $\mathcal P(\omega')$.
For the filter function with the box shape (\ref{filter_box}), it is further simplified as
\begin{align}
	C_\ell(\omega) = \frac{\omega}{\Delta\omega} \frac{H(z)}{(1+z)r^2(z)} P_\rho\left(k=\frac{\ell}{r(z)}; z\right),
    \label{Cl_dec_app}
\end{align}
where $1+z=m/(2\omega)$ and assumed $\Delta\omega \lesssim \omega$.
We can clearly see the enhancement factor $\omega/\Delta\omega$.

At first sight, the enhancement of $\omega/\Delta\omega$ may look strange, since $\Delta\omega$ is purely determined by the detector properties and $C_\ell$ diverges for $\Delta\omega\to 0$. 
Such a dependence appears only in the case of line gammas at the production.
If one neglects the dark matter velocity dispersion and fix the observation photon energy $\omega$, the redshift at which the photons have been produced is uniquely determined.
Since the angular power spectrum sees the correlation between photons with fixed energy from different directions, it means that it is looking at the spatial correlation of the matter density at the same redshift.
According to Eq.~(\ref{Limber}), the correlation of the matter density is strongly peaked at the same redshift. In the lowest order approximation it has a delta function like peak, leading to the divergence for  $\Delta\omega\to 0$. 
Therefore, the ultimate limit of the applicability of (\ref{Cl_dec_app}) may come from the breakdown of the Limber approximation (\ref{Limber}). In reality, the broadening of the line gamma spectrum might be more important. In our analysis in the following, we only consider $\omega/\Delta\omega \lesssim 10^{2}$ so that these effects are not very important. 

The expression for $C_\ell$ (\ref{Cl_dec}) or its approximate form (\ref{Cl_dec_app}) has no explicit dependence on the dark matter model parameters and only determined by the astrophysical quantities, since we defined $C_\ell$ as the ratio between the angular correlations and the squared mean intensity.
The dark matter information implicitly enters only through the redshift $1+z=m/(2\omega)$.
Conversely, once $C_\ell$ is calculated for various redshift $z$, it is applicable for any dark matter model that satisfies $1+z=m/(2\omega)$.
If one wants to compare the angular power spectrum for given dark matter model with real observational data, one may just multiply our $C_\ell$ with the mean intensity squared, which is much easier than calculating the angular power spectrum directly.

\begin{figure}\centering
\includegraphics[width=.48\textwidth]{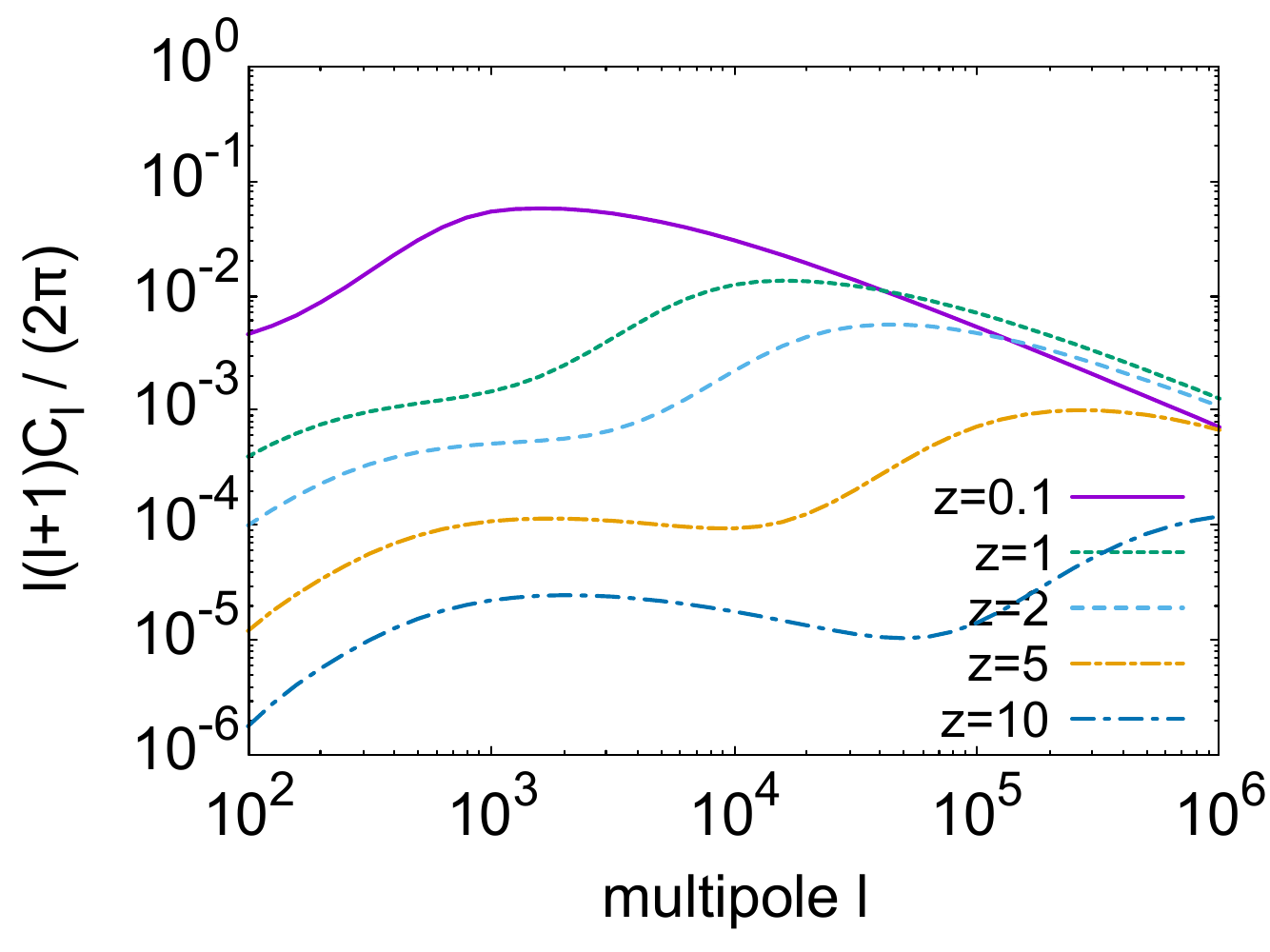}
\includegraphics[width=.48\textwidth]{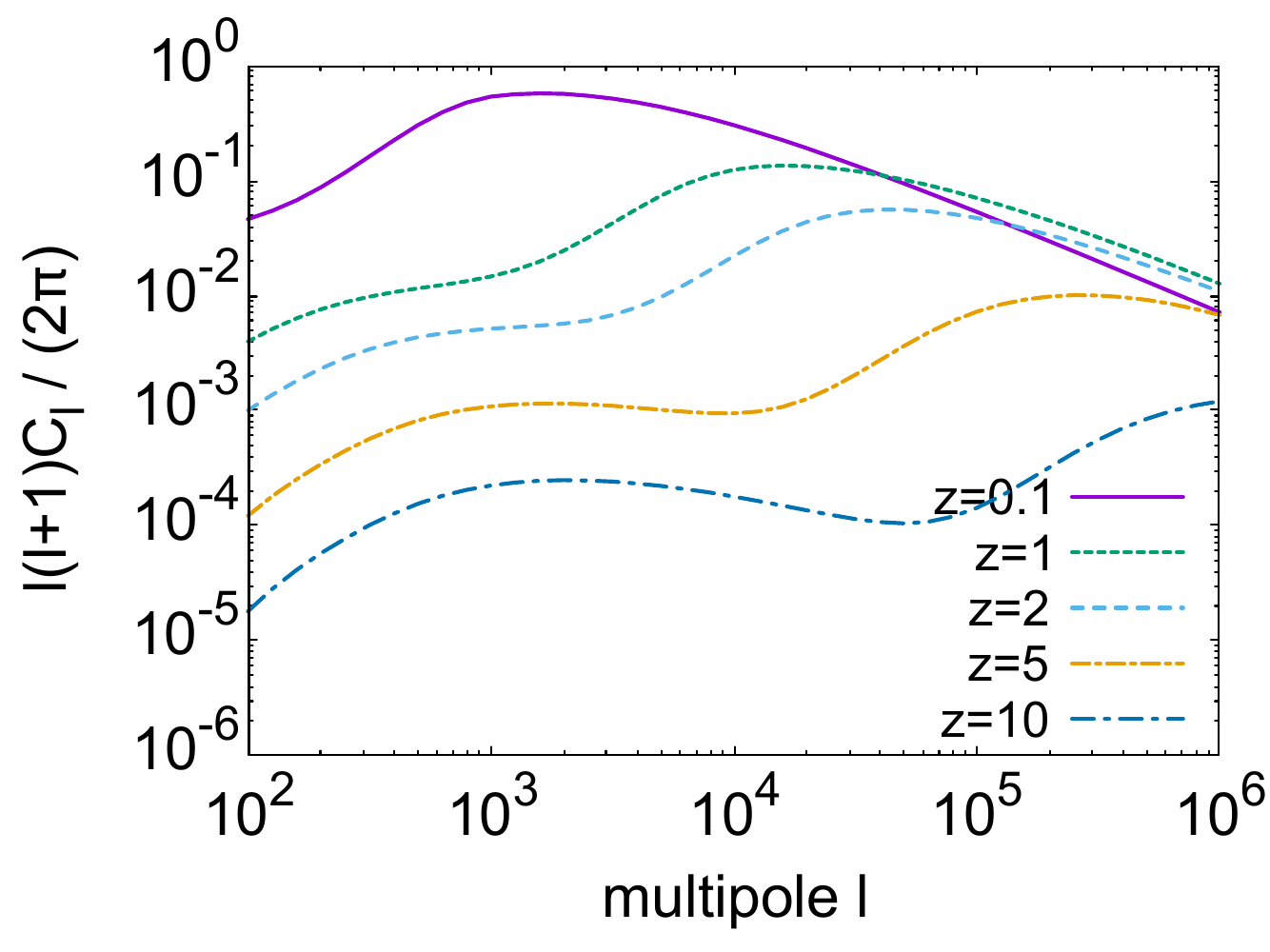}
\caption{The dimensionless angular power spectrum of the cosmic background photon from dark matter decay for various redshift. The horizontal axis is the multipole $\ell$ and the vertical axis is $\ell(\ell+1)C_\ell/(2\pi)$. We have taken $\omega/\Delta\omega=1$ in the left panel and $\omega/\Delta\omega=10$ in the right panel where $\omega$ is the observation photon energy and $\Delta\omega$ is the energy resolution of the detector. We are assuming dark matter decaying into line photons and hence the redshift at which the photons are produced is uniquely determined, once the dark matter mass and the observation photon energy is fixed.}
\label{fig:Cl_dec}
\end{figure}

Fig.~\ref{fig:Cl_dec} shows the dimensionless angular power spectrum $\ell(\ell+1)C_\ell/(2\pi)$ from dark matter decay for various redshift. We have taken $\omega/\Delta\omega=1$ in the left panel and $\omega/\Delta\omega=10$ in the right panel.
The left peak (large angular scale peak) corresponds to the 2-halo contribution and the right peak (small angular scale peak) corresponds to the 1-halo contribution.

\subsection{Annihilating dark matter}

The evaluation of photon flux and its angular power spectrum for the annihilating dark matter case is more involved than the case of decaying dark matter.
One important difference is that the dark matter annihilation rate is proportional to the square density and hence we need to evaluate the mean variance $\left<\rho_m^2\right>$ and also the spatial correlation of square density rather than the density itself.
It requires additional considerations and techniques as we will see below.

\subsubsection{Isotropic flux}

The evaluation of isotropic  gamma-ray flux from dark matter annihilation has been done in Ref.~\cite{Ullio:2002pj}. 
Below we mostly follow Refs.~\cite{Ullio:2002pj,Kawasaki:2009nr}.

\begin{figure}\centering
\includegraphics[width=.48\textwidth]{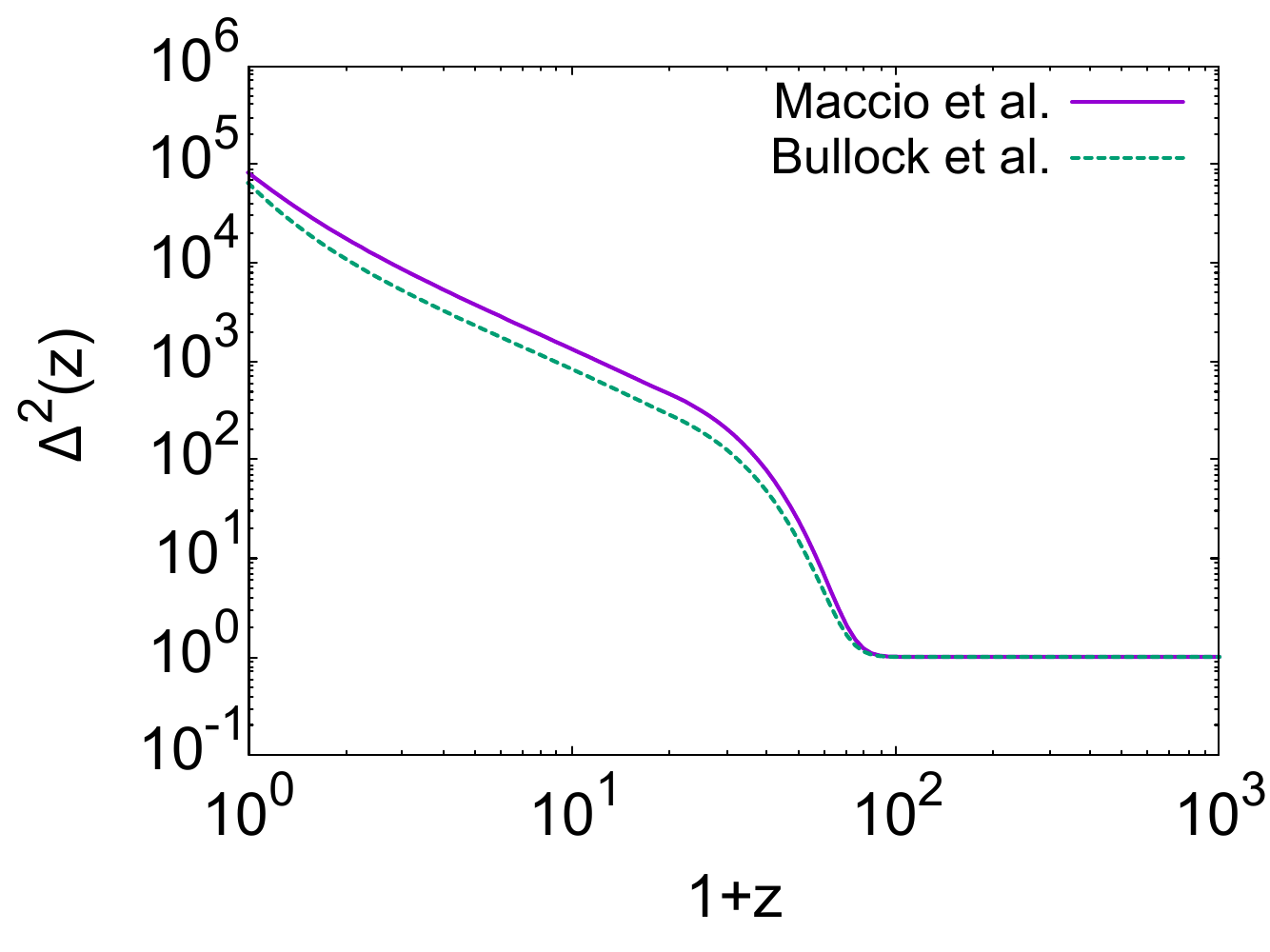}
\includegraphics[width=.48\textwidth]{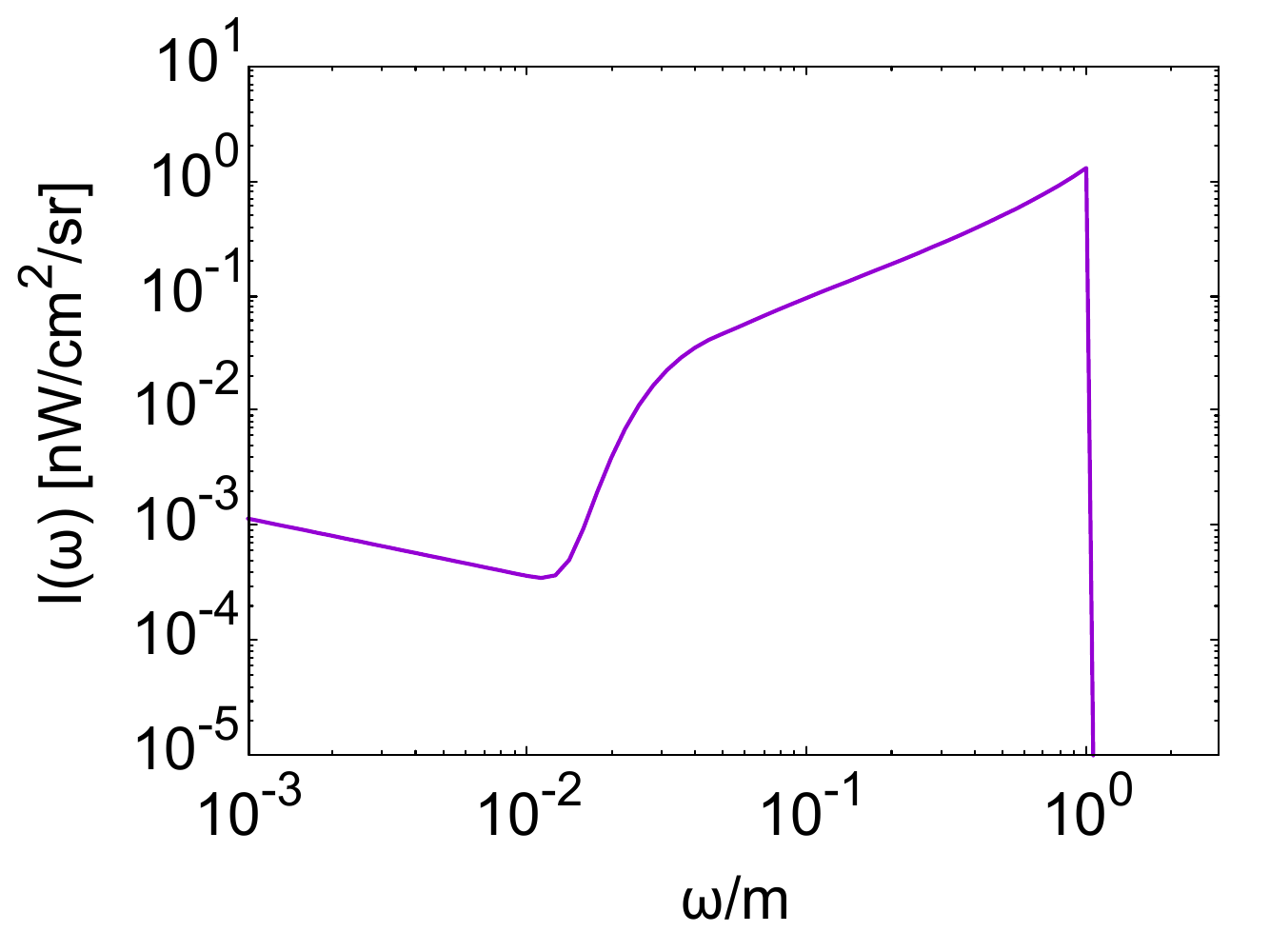}
\caption{(Left) The enhancement factor $\Delta^2(z)$ as a function of $1+z$, calculated with two prescriptions for the concentration parameter (see Eq.~(\ref{cvir_Bullock}) and (\ref{cvir_Maccio})).
(Right) The mean intensity of cosmic background photon from dark matter annihilating into line photons. We have taken $N_\gamma\left<\sigma v\right>/m=10^{-38}\,{\rm cm^2/eV}$.}
\label{fig:Delta2}
\end{figure}

Similarly to the case of decaying dark matter, the mean photon flux averaged over the observation bandwidth is given by
\begin{align}
	\overline I(\omega) = \int d\omega' \left(\omega'\right)^q \mathcal P(\omega') \int_0^\infty dr W(r,\omega'') \Delta^2(z),
	\label{Imean_ann}
\end{align}
where $\omega''=(1+z)\omega'$,
\begin{align}
	W(r,\omega'')= \sum_F \frac{1}{8\pi} \frac{\langle{\rho_{\rm DM}}(z,\vec x)\rangle^2}{m^2 (1+z)^3}\, \langle\sigma v\rangle_F\, \frac{d\mathcal N^{(F)}_\gamma}{d\omega''},
\end{align}
with $\langle\sigma v\rangle_F$ being the dark matter annihilation cross section\footnote{
    In this paper we only consider the case of $\left<\sigma v\right> = {\rm const}.$ } into an arbitrary final state $F$ and we have defined
\begin{align}
	\Delta^2(z) \equiv \frac{\langle\rho^2_{\rm DM}(z,\vec x)\rangle}{\langle\rho_{\rm DM}(z,\vec x) \rangle^2}
	= \left<\delta_\rho^2(\vec x;z)\right> + 1.
\end{align}
What is nontrivial compared with the case of decay is $\langle\rho^2_{\rm DM}(z)\rangle \neq \langle\rho_{\rm DM}(z)\rangle^2$. Actually, due to the nonlinear structure formation $\Delta^2(z)$ is orders of magnitude larger that the latter at late epoch. 
We obtain $\Delta^2(z)$ by taking $\vec x_1=\vec x_2$ in Eq.~(\ref{deltarho_2p}) in Appendix~\ref{app:mat}:
\begin{align}
	\Delta^2(z) &= \frac{1}{\overline \rho_{m0}^2} \int dM\,M^2\frac{dn_h(z)}{dM} \int\frac{d^3k}{(2\pi)^3}\left|u_M(k;z)\right|^2 + 1 \nonumber\\
    &=\frac{(1+z)^3}{\overline \rho_{m0}^2} \int dM\,\frac{dn_h(z)}{dM} \int dr\,4\pi r^2 \rho_h^2(r;M;z)  + 1
    \label{Delta2}
\end{align}
It is not difficult to show that this quantity is the same as $\Delta^2(z)$ in Refs.~\cite{Ullio:2002pj,Kawasaki:2009nr}. For the Navarro-Frenk-White density profile (\ref{NFW}), it is calculated as
\begin{align}
	\Delta^2(z) &= \frac{\Delta_{\rm vir}(z)}{9\overline \rho_{m0}} \int dM\,M\frac{dn_h(M;z)}{dM} \frac{c_{\rm vir}^3(M;z)}{\mathcal F^2(c_{\rm vir}(M;z))}\left(1-\frac{1}{(1+c_{\rm vir}(M;z))^3}\right)+1.
    \label{Delta2}
\end{align}
See Appendix~\ref{app:density} for the definition of $\Delta_{\rm vir}(z)$, $c_{\rm vir}(M; z)$ and $\mathcal F(c_{\rm vir}(M;z))$.
The resulting value of $\Delta^2(z)$ is shown in the left panel of Fig.~\ref{fig:Delta2}. Two lines correspond to two prescriptions for the concentration parameter $c_{\rm vir}(M;z)$ (see Eq.~(\ref{cvir_Bullock}) and (\ref{cvir_Maccio})).

Now let us suppose that the dark matter annihilates directly into photons. Similarly to the decaying dark matter case, we consider final states $F=2\gamma$ or $F=\gamma+X$. Then the multiplicity function is given by
\begin{align}
	\frac{d\mathcal N_\gamma}{d\omega''} = \frac{N_\gamma}{2} \delta\left(\omega'' - m\right),
	\label{dNdomega_ann}
\end{align}
where $N_\gamma=2$ or $1$. Substituting this into (\ref{Imean_ann}), we obtain
\begin{align}
	\overline I(\omega) = \int d\omega' \left(\omega'\right)^{q-4} \mathcal P(\omega') \frac{m\rho^2_{\rm DM,0} \Delta^2(z)}{16\pi} \left.\frac{N_\gamma\left<\sigma v\right>}{H(z)}\right|_{1+z=\frac{m}{\omega'}}.
\end{align}
In the narrow band limit $\mathcal P(\omega')=\delta(\omega-\omega')$, the isotropic flux becomes
\begin{align}
	&\overline I(\omega) = \frac{\omega^{q-4}\,m\,\rho^2_{\rm DM,0} \Delta^2(z)}{16\pi} \left.\frac{N_\gamma\left<\sigma v\right>}{H(z)}\right|_{1+z=\frac{m}{\omega}} \nonumber \\
    &~~\simeq 6.6\,[{\rm nW/m^2/sr}] \left(\frac{m}{\omega}\right)^2 \left(\frac{N_\gamma\left<\sigma v\right>/m}{10^{-31}\,{\rm cm^3/s/eV}}\right) \left(\frac{\Delta^2\left(z\right)}{10^5}\right)
    \left[\Omega_{\Lambda 0} + \Omega_{m0}\left(\frac{m}{\omega}\right)^3\right]^{-1/2},
\end{align}
where we assumed $q=2$ in the second line.
The resulting photon flux is shown in the right panel of Fig.~\ref{fig:Delta2}. It is seen that it has nontrivial frequency dependence, reflecting the redshift dependence of the enhancement factor $\Delta^2(z)$.
Note again that the use of $\mathcal P(\omega')=\delta(\omega-\omega')$ is not justified when evaluating the angular power spectrum, as seen below.

\subsubsection{Angular power spectrum}

Next let us evaluate the angular power spectrum from dark matter annihilation~\cite{Ando:2005xg,Ando:2013ff,Fornengo:2013rga}.
The fluctuation of the photon intensity is given by
\begin{align}
	\delta_I(\omega,\hat\Omega) &= \frac{1}{\overline I(\omega)} \int d\omega' (\omega')^{q}\mathcal P(\omega') \int dr W(r,\omega'') \Delta^2(z)\delta_{\rho^2}(r,\hat\Omega),
\end{align}
where 
\begin{align}
	\delta_{\rho^2}(z,\vec x) = \frac{\rho^2(z,\vec x) - \langle\rho^2(z,\vec x)\rangle}{\langle\rho^2(z,\vec x)\rangle} = \frac{\rho^2(z,\vec x)}{\Delta^2(z)\bar\rho(z)^2}-1.
    \label{delta_rho2}
\end{align}
The angular power spectrum is given in a similar fashion to derive Eq.~(\ref{Cl_general}) as
\begin{align}
    C_\ell(\omega)&=\frac{1}{\overline I^2(\omega)}\int d\omega_1'(\omega_1')^{q}\mathcal P(\omega_1') \int d\omega_2'(\omega_2')^{q}\mathcal P(\omega_2') \nonumber \\
    &~~~~~\times \int \frac{dr_1}{r_1^2} W(r_1,\omega_1'')W(r_1,\omega_2'')\Delta^4(z_1)P_{\rho^2}\left(k=\frac{\ell}{r_1}; z_1\right).
    \label{Cell_ann}
\end{align}
The power spectrum $P_{\delta^2}(k;z)$ is calculated in Appendix~\ref{app:4point}. 
Note that the explicit $\Delta^4(z_1)$ dependence in the second line of (\ref{Cell_ann}) is canceled out due to the $\Delta^{-4}(z_1)$ dependence of $P_{\rho^2}\left(k=\frac{\ell}{r_1}; z_1\right)$.
Still one has $\Delta(z_1)^{-4}$ dependence from $\overline I^{-2}$ (see Eq.~(\ref{Imean_ann})).

For the line gamma spectrum (\ref{dNdomega_ann}), we obtain
\begin{align}
	C_\ell(\omega) &= \frac{1}{\overline I^2(\omega)}\int dz \left(\frac{(\omega')^q \mathcal P(\omega')}{1+z}\frac{\rho^2_{\rm DM,0} N_\gamma \langle\sigma v\rangle (1+z)^3}{16\pi m^2} \right)^2 \frac{ \Delta^4(z)}{H(z) r^2(z)}P_{\rho^2}\left(k=\frac{\ell}{r(z)}; z\right) \nonumber\\
    &=\frac{\int dz \left(\frac{(\omega')^q \mathcal P(\omega')(1+z)^2\Delta^2(z)}{H(z)}\right)^2 \frac{H(z)}{r^2(z)} P_{\rho^2}\left(k=\frac{\ell}{r(z)}; z\right)}{\left[\int dz \frac{(\omega')^q \mathcal P(\omega')(1+z)^2\Delta^2(z)}{H(z)}  \right]^2}.
\end{align}
For the filter function with the box shape (\ref{filter_box}), it is further simplified as
\begin{align}
	C_\ell(\omega) = \frac{\omega}{\Delta\omega} \frac{H(z)}{(1+z)r^2(z)} P_{\rho^2}\left(k=\frac{\ell}{r(z)}; z\right),
\end{align}
where $1+z=m/\omega$ and assumed $\Delta\omega \lesssim \omega$.
This is exactly the same form as the case of decaying dark matter after replacing $P_{\rho}\to P_{\rho^2}$. 
Again, we can clearly see the enhancement factor $\omega/\Delta\omega$.

\begin{figure}\centering
\includegraphics[width=.48\textwidth]{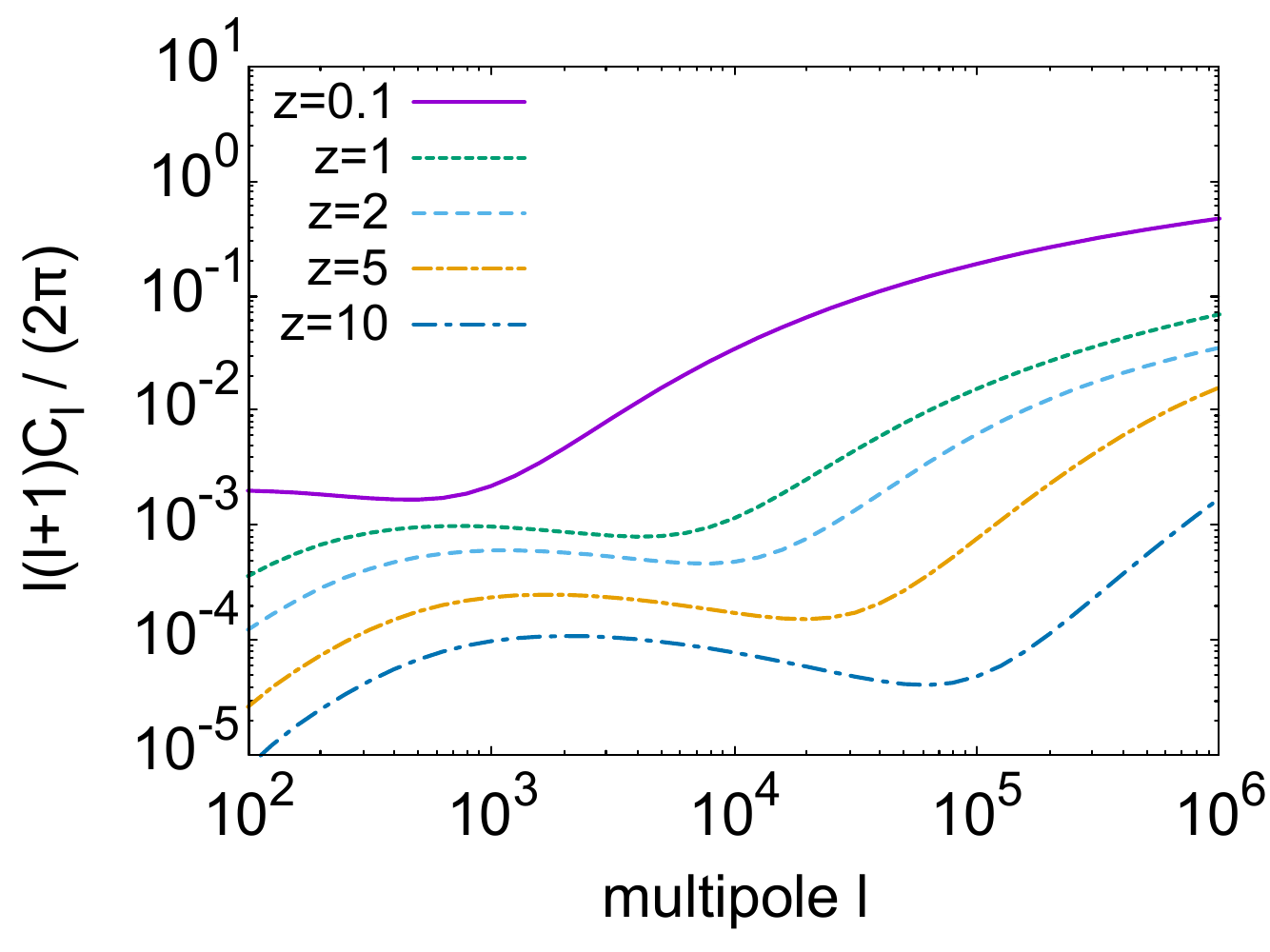}
\includegraphics[width=.48\textwidth]{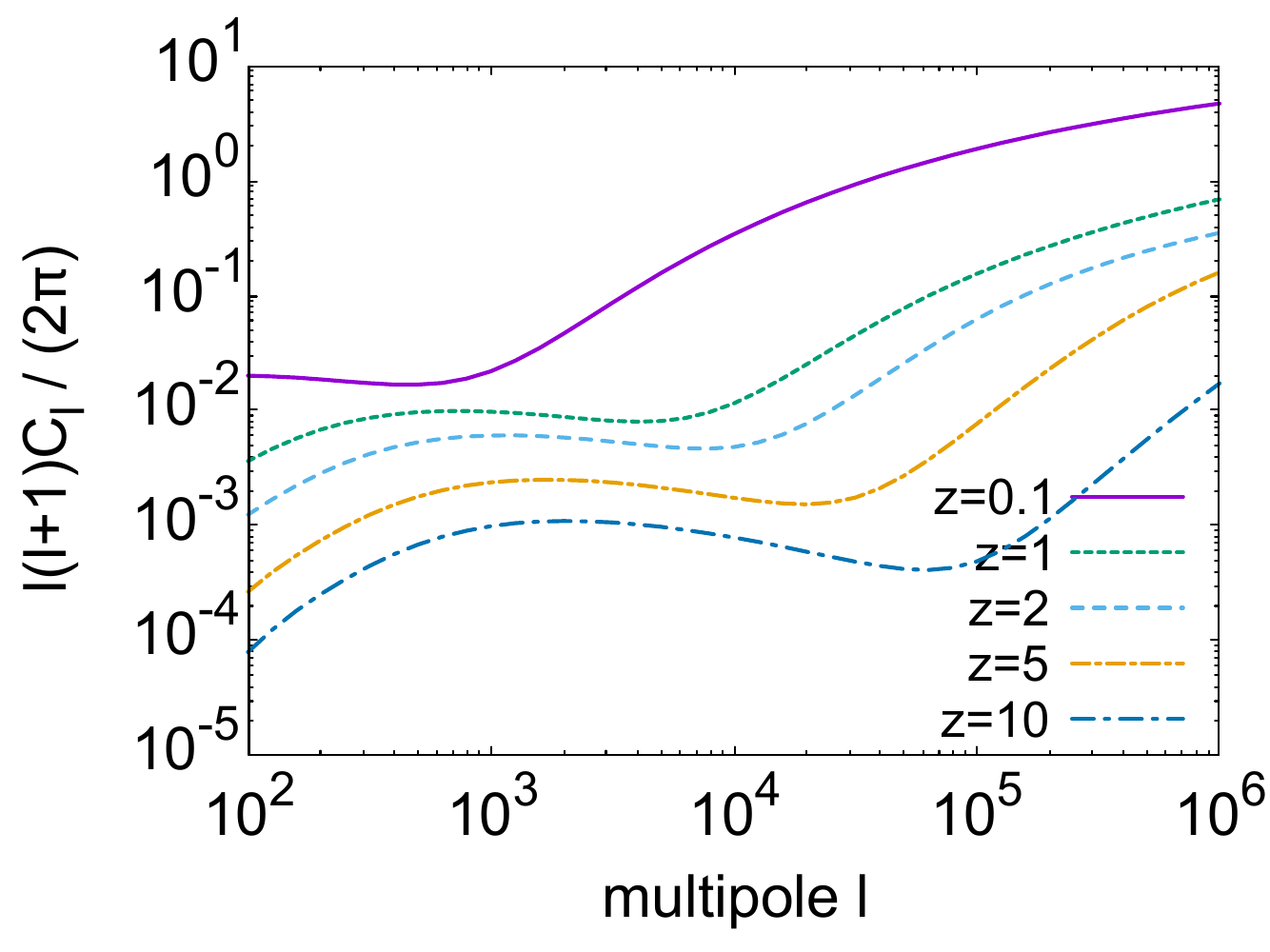}
\caption{The dimensionless angular power spectrum of the cosmic background photon from dark matter annihilation for various redshift. The horizontal axis is the multipole $\ell$ and the vertical axis is $\ell(\ell+1)C_\ell/(2\pi)$. We have taken $\omega/\Delta\omega=1$ in the left panel and $\omega/\Delta\omega=10$ in the right panel. We are assuming dark matter annihilating into line photons.}
\label{fig:Cl_ann}
\end{figure}

Fig.~\ref{fig:Cl_ann} shows the dimensionless angular power spectrum $\ell(\ell+1)C_\ell/(2\pi)$ from dark matter annihilation for various redshift. We have taken $\omega/\Delta\omega=1$ in the left panel and $\omega/\Delta\omega=10$ in the right panel.

\section{Application}
\label{sec:app}

In this section we derive constraints on the dark matter decay rate and annihilation cross section into line photons based on the formalism developed so far.
We make use of several observational data of the angular power spectrum of the background photons.

\paragraph{From radio to ultraviolet observations}

\begin{figure}\centering
\includegraphics[width=.9\textwidth]{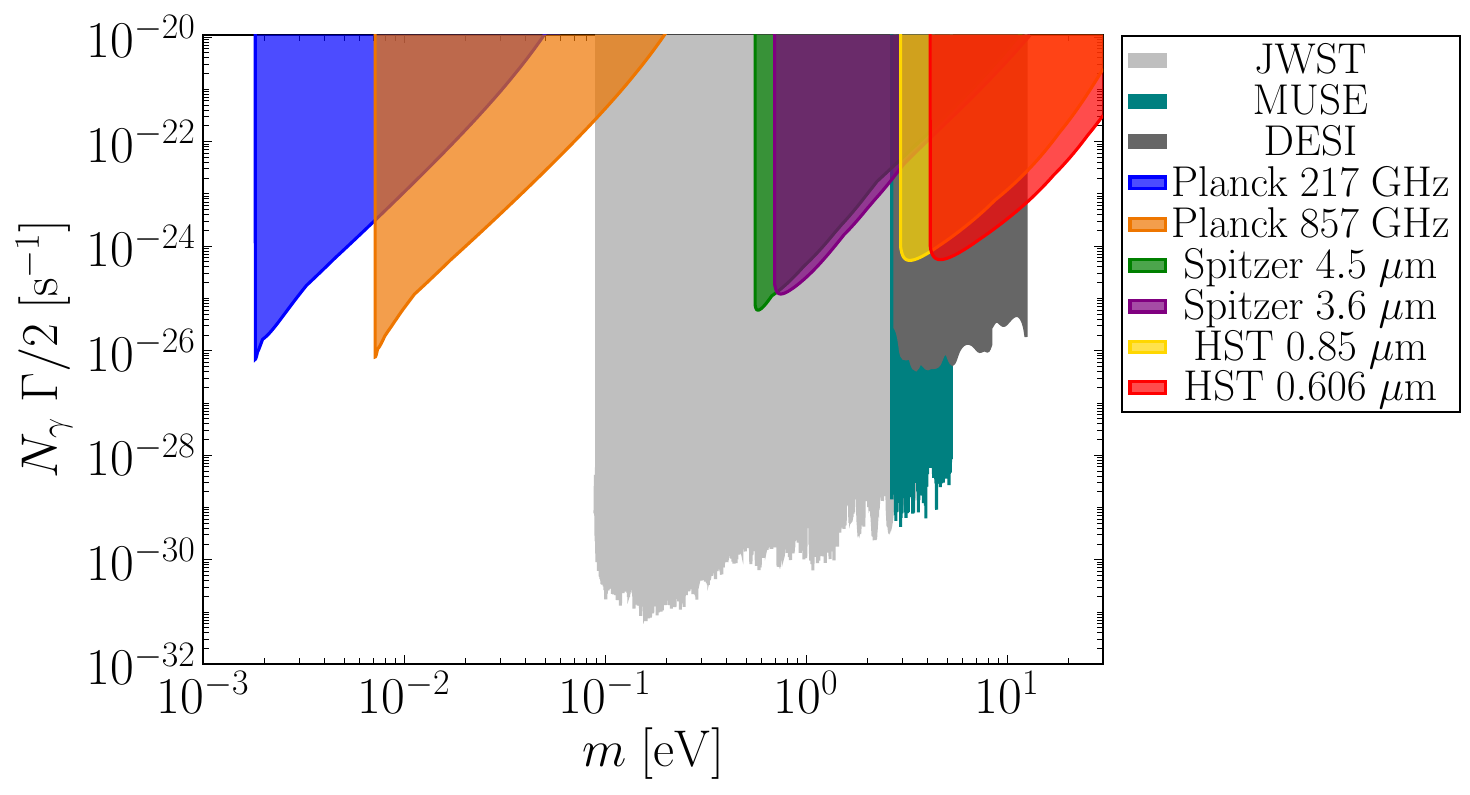}
\caption{Constraints on the dark matter decay rate into photons from the angular power spectrum measured by Planck, IRAC/Spitzer, and HST with several observation frequencies.
Constraints from spectroscopic searches for line photons by the DESI, JWST and MUSE telescopes are also shown.}
\label{fig:decay_constraint}
\end{figure}
\begin{figure}\centering
\includegraphics[width=.9\textwidth]{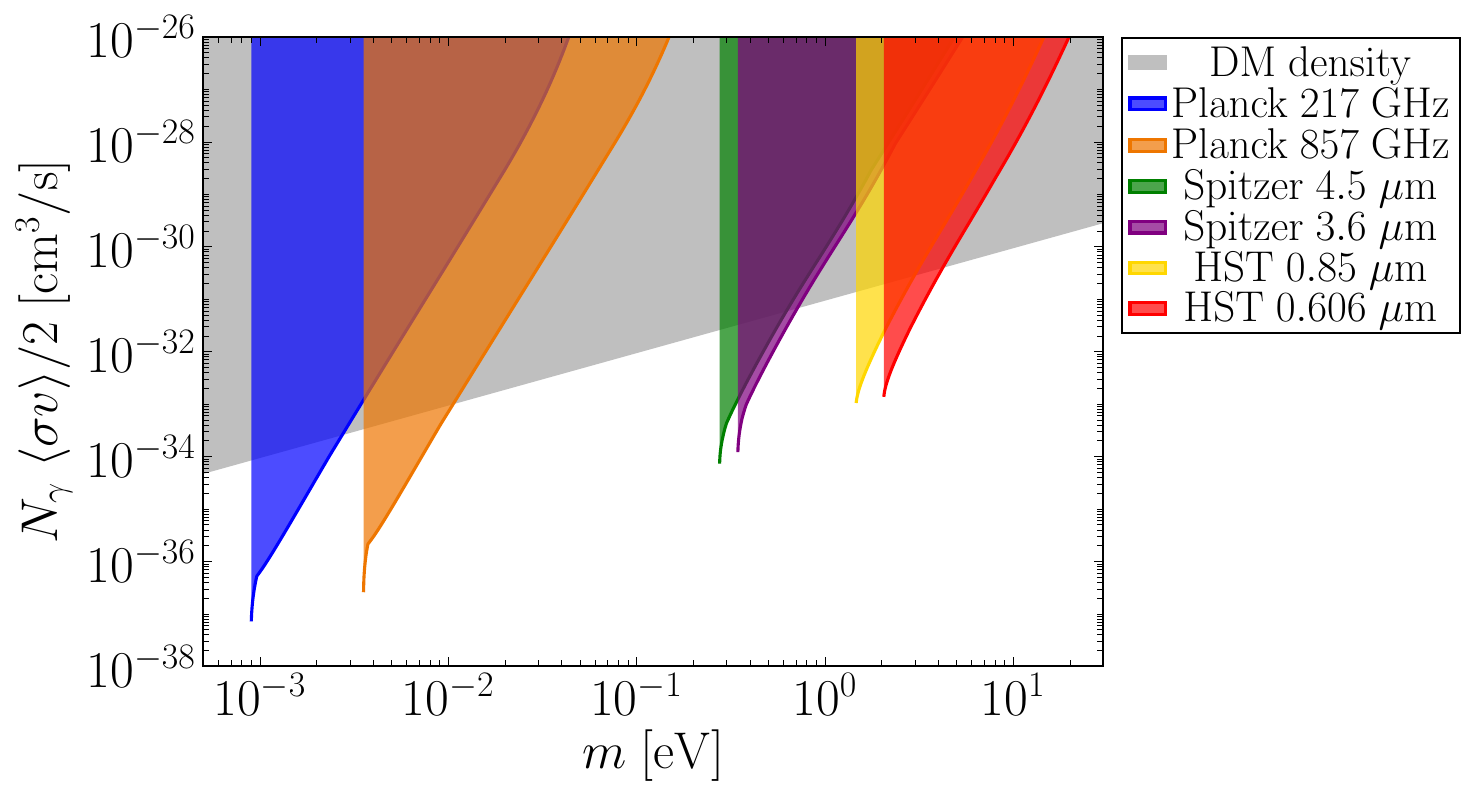}
\caption{Constraints on the dark matter annihilation cross section into photons from the angular power spectrum measured by Planck, IRAC/Spitzer, and HST with several observation frequencies.}
\label{fig:ann_constraint}
\end{figure}

First, let us consider the frequency range from radio/infrared to optical/ultraviolet. 
We compare our theoretical predictions from dark matter decay/annihilation with the Planck satellite data~\cite{Planck:2011ivn} for radio/infrared frequency, the IRAC instrument of the Spitzer satellite~\cite{Kashlinsky:2012zz} for infrared frequency, and the Hubble Space Telescope (HST) data~\cite{Kashlinsky:2018mnu} for optical/ultraviolet frequency.\footnote{
    Ref.~\cite{Kashlinsky:2018mnu} plots the angular power spectrum in terms of $\frac{2\pi}{q}\,{\rm [arcsec]}$. It is converted to $\ell$ through $\ell = \left(\frac{2\pi}{q}\,{\rm [arcsec]}\right)^{-1}\times 3600\times 360$.}
Fig.~\ref{fig:decay_constraint} shows the resulting constraint on the dark matter decay rate and Fig.~\ref{fig:ann_constraint} shows constraints on the annihilation cross section.
We used $\omega/\Delta\omega= 3.03$ $(3.33)$ for Planck $217$ $(857)\,\mathrm{GHz}$~\cite{PlanckHFICoreTeam:2011az},\footnote{As written in Ref.~\cite{Planck:2015mrs}, Planck $217\,\mathrm{GHz}$ result is a combination of measurements from 217P channel and 217 channel. We choose 217 channel's bandwidth for $\Delta\omega$ since wider bandwidth outputs weaker constraints.} $4.35$ $(4.76)$ for IRAC/Spitzer $4.5$ $(3.6)\,\mu \mathrm{m}$~\cite{IRAC:2004rru}, $7.13$ $(3.74)$ for HST $0.85$ $(0.606)\,\mu\mathrm{m}$~\cite{stsci_acs_filters}.
Together also shown are existing constraints on decaying dark matter obtained by spectroscopic searches for line photons from galaxies or blank sky with the DESI~\cite{Wang:2023imi}, JWST~\cite{Pinetti:2025owq} and MUSE~\cite{Todarello:2023hdk} telescopes.\footnote{
    For these existing constraints, including X-ray constraints in the next subsection (``XMM-Newton, NuSTAR and X-ray (2011)''), we used the data file in \cite{AxionLimits}.}
As for the constraint from HST, our results are consistent with previous studies~\cite{Nakayama:2022jza,Carenza:2023qxh}.
For the dark matter annihilation case, we have also shown a constraint $\left<\sigma v\right>/m \lesssim 10^{-31}\,{\rm cm^3/s/eV}$ so that the pair annihilation does not significantly reduce the dark matter density below the redshift $z\sim 10^6$.
We found new constraints from the IRAC/Spitzer and Planck data. The IRAC/Spitzer constraint is weaker than the spectroscopic searches, while the Planck gives meaningful constraints.\footnote{
    If we assume an axion-like particle decaying into two photons, the new constraint regions by the Planck are already excluded by the stellar constraints~\cite{Raffelt:1996wa}.
}

\paragraph{X-ray observations}

\begin{figure}\centering
\includegraphics[width=.9\textwidth]{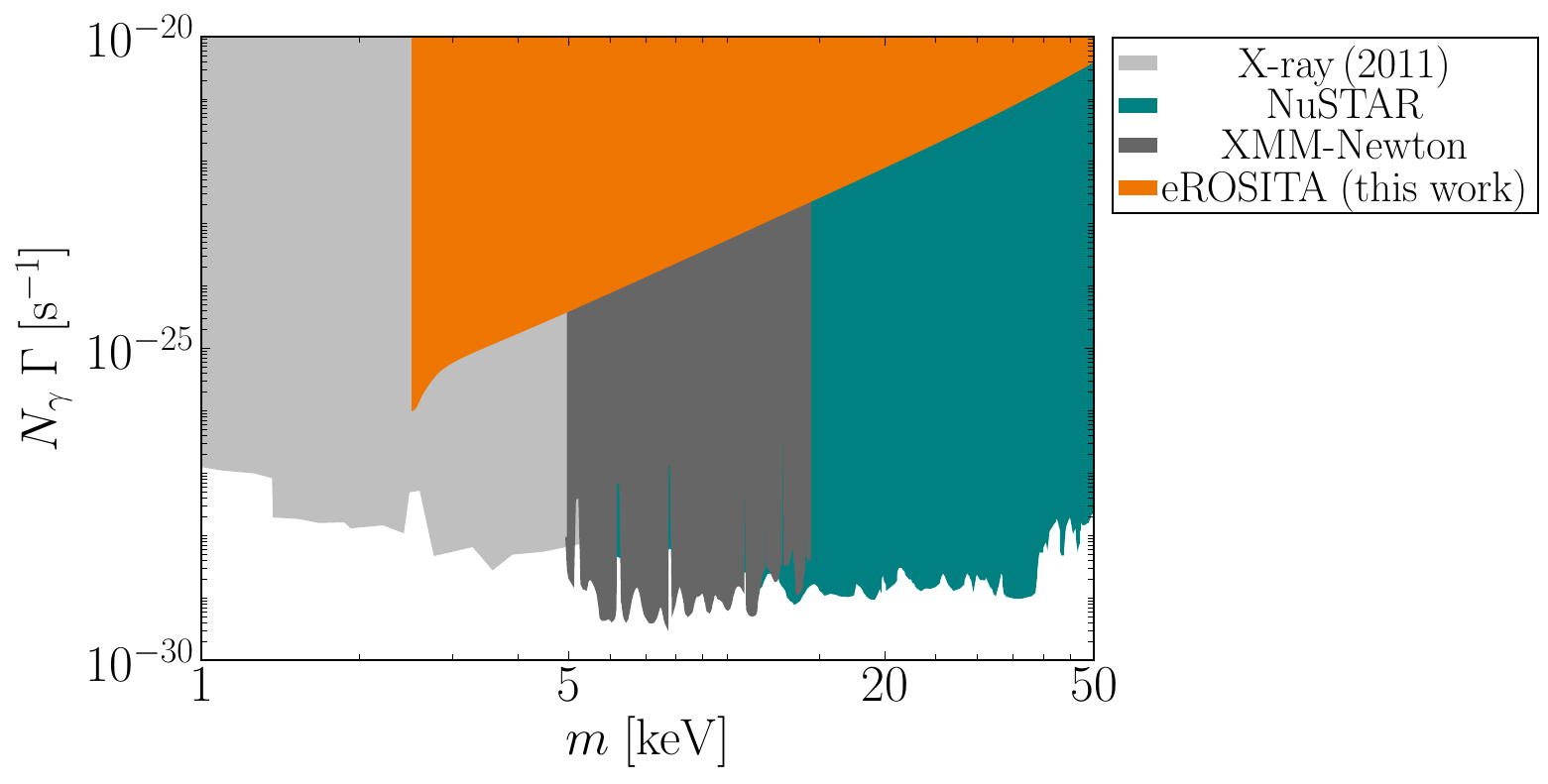}
\caption{Constraints on the dark matter decay rate into photons from the angular power spectrum measured by the eROSITA all sky survey.
Constraints from spectroscopic searches for line photons by the XMM-Newton, NuSTAR and combined constraint as of 2011 are also shown.}
\label{fig:decay_constraint_X}
\end{figure}
\begin{figure}\centering
\includegraphics[width=.9\textwidth]{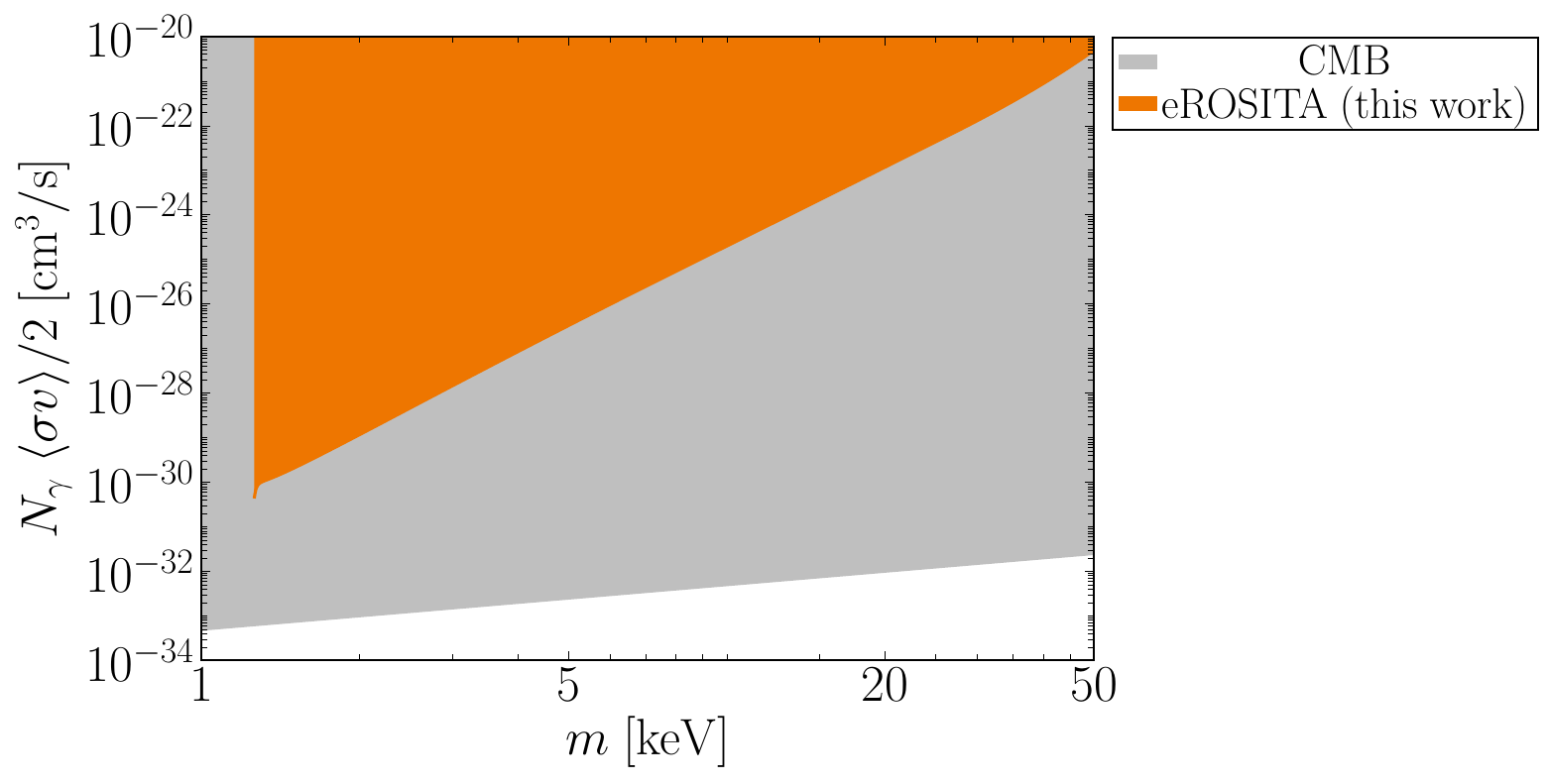}
\caption{Constraints on the dark matter annihilation cross section into photons from the angular power spectrum measured by the eROSITA all sky survey.}
\label{fig:ann_constraint_X}
\end{figure}

Next let us see constraints from X-ray observations.
Fig.~\ref{fig:decay_constraint_X} shows a constraint on the dark matter decay rate, and
Fig.~\ref{fig:ann_constraint_X} shows a constraint on the annihilation cross section,
both from the angular power spectrum measurement by the eROSITA all sky survey~\cite{Lau:2024xrd}.
We take $\omega=1.25\,\mathrm{keV}$, $\omega/\Delta\omega= 0.83$ and the averaged area of the instrument as $0.21\,\mathrm{m^2}$ to convert the flux per unit area to the number of detected particles in the detector.\footnote{We use the information in TM0 on-axis data from \cite{eROSITA_DR1_ARFRMF}.}
In Fig.~\ref{fig:decay_constraint_X}, we also show constraints from the X-ray line searches by the XMM-Newton~\cite{Foster:2021ngm}, NuSTAR~\cite{Perez:2016tcq,Ng:2019gch,Roach:2022lgo} and also X-ray constraints as of 2011 summarized in Ref.~\cite{Cadamuro:2011fd}.
The energy injection by dark matter annihilation around the recombination epoch changes the ionization fraction of the hydrogen atom, affecting the Cosmic Microwave Background (CMB) anisotropy~\cite{Padmanabhan:2005es,Slatyer:2009yq,Kanzaki:2009hf,Finkbeiner:2011dx,Slatyer:2012yq,Kawasaki:2015peu,Planck:2018vyg,Kawasaki:2021etm}. 
Constraint from the CMB observation is shown in Fig.~\ref{fig:ann_constraint_X}.
It is seen that the CMB constraint is much stronger than the constraint from X-ray angular power spectrum.

\section{Conclusions}
\label{sec:conc}

Founding dark matter signals through cosmic rays is a long-standing issue. 
The energy and composition of cosmic rays depend on the dark matter model, and various observations may have a potential to find dark matter signals.
In this paper we concentrated on the angular power spectrum of photons predicted by dark matter decay or annihilation. 
Since dark matter distribution is inhomogeneous in the universe, the cosmological flux of photons produced by dark matter decay/annihilation is also inhomogeneous, resulting in the angular power spectrum.
This has already been pointed out in several papers in the past, but relatively less investigation has been done.
In particular, when the dark matter decay/annihilates into line photons, a naive calculation breaks down due to the artificial divergence and we need to take into account either the detector energy resolution or some intrinsic line broadening effects.
If the dark matter is lighter than the electron mass scale, the only possible decay/annihilation modes may be photons or neutrinos, which necessarily have line spectrum.
For the first time, we formulated the calculation of angular power spectrum arising from dark matter annihilation into line photons.
We note that, for calculating the angular power spectrum, various astrophysical quantities need to be evaluated, such as halo mass function, halo density profile and its Fourier transformation, concentration parameter, nonlinear power spectrum of the density or density square.
These quantities are not often presented in detail in the literature, so in this paper we tried to present them in detail as far as we can, which we believe are beneficial for readers.
We also emphasize that our expression for the \textit{dimensionless} angular power spectrum $C_\ell$ is independent of the dark matter model parameters. 
Thus one can estimate the dimensionful angular power spectrum just by multiplying the dimensionless power spectrum with the mean intensity from dark matter decay/annihilation. 
It should also be convenient for readers.

We applied our formalism to calculate the angular power spectrum in order to compare it with observations with various telescopes from radio to X-rays. 
Results are summarized in Figs.~\ref{fig:decay_constraint}, \ref{fig:decay_constraint_X} and \ref{fig:ann_constraint_X}.
In the frequency range where spectroscopic surveys are available, constraints from the angular power spectrum are typically weaker.
However, since the angular power spectrum we calculated is of cosmological origin, wide mass range is covered by a single frequency observation and hence it can cover the dark matter mass range outside the spectroscopic surveys.
Moreover, the sensitivity will increase by narrowing the detector bandwidth, i.e., increasing $\omega/\Delta\omega$. 
If line photon signals will be found, the analysis of angular power spectrum may confirm whether it is dark matter origin or not.

Finally we mention that our formalism may have broader applications.
For example, cosmological dark matter decay/annihilation may produce neutrinos, which should also have fluctuations similar to the photon.
Recently it is pointed out that gravitons may also be produced by dark matter decay/annihilation~\cite{Ema:2021fdz,Strumia:2025dfn}.
Although the total intensity for them is highly dependent on dark matter models, the (dimensionless) angular power spectrum is independent of the decay products, as far as they are relativistic and not scattered/absorbed by Galactic/intergalactic medium.

\section*{Acknowledgment}

This work was supported by World Premier International Research Center Initiative (WPI), MEXT, Japan.
This work was also supported by JSPS KAKENHI (Grant Number 24K07010 [KN], 26K00695 [KN], 26H00403 [KN]).

\appendix
\section{Nonlinear matter power spectrum} \label{app:mat}

As discussed in the main text, in order to estimate the angular power spectrum of the cosmic photon background from dark matter annihilation, we need to evaluate the nonlinear power spectrum of the matter density $\rho_m(\vec x,z)$ for the case of decaying dark matter, and that of the square matter density $\rho_m^2(\vec x,z)$ for the case of annihilating dark matter.
In this Appendix we summarize methods to evaluate them.

\subsection{Power spectrum for matter density}
\label{app:2point}

Below we basically follow Refs.~\cite{1991ApJ...381..349S,Cooray:2002dia}.
The matter density field $\rho_m(\vec x)$ is rewritten as
\begin{align}
	\rho_m(\vec x,z) &= \sum_i \rho_h\left(\frac{\vec x-\vec x_i}{1+z}; z; M_i\right) \equiv \sum_i (1+z)^3M_i u_{M_i}(\vec x-\vec x_i; z) \nonumber\\
	&=\int dM \int d^3x' \left[\sum_i \delta(\vec x'-\vec x_i) \delta(M-M_i)\right] (1+z)^3M u_M(\vec x-\vec x'; z),
\end{align}
where $i$ labels halo whose center position is $\vec x_i$, and $\rho_h(\vec x-\vec x_i; M_i)$ denotes the density profile of the halo with the halo mass $M_i$.
The normalized density profile satisfies $\int d^3x\,u_M(\vec x-\vec x_i) = 1$.
We note that the halo mass function is given by
\begin{align}
	\frac{dn_h(z)}{dM} = \left<  \sum_i \delta(\vec x'-\vec x_i) \delta(M-M_i) \right>.
\end{align}
Note that the halo mass function is defined as a comoving number density.
From this we immediately see
\begin{align}
	\overline\rho_m(z) \equiv \langle\rho_m(\vec x, z)\rangle = (1+z)^{3}\int dM\,M\frac{dn_h(z)}{dM}.
\end{align}
Then let us consider the 2 point correlation of the density fluctuation field $\delta_\rho(\vec x,z)\equiv \left(\rho_m(\vec x, z)-\overline\rho_m(z)\right) / \overline\rho_m(z)$:
\begin{align}
	&\left<\delta_\rho(\vec x_1,z)\delta_\rho(\vec x_2,z)\right> +1= \frac{1}{\overline\rho_0^2}\int dM_1\int dM_2\int d^3x_1'\int d^3x_2'
	\, u_{M_1}(\vec x_1-\vec x_1';z)u_{M_2}(\vec x_2-\vec x_2';z) \nonumber \\
	&~~~~~~\times M_1M_2\left< \sum_i\sum_j \delta(\vec x_1'-\vec x_i) \delta(\vec x_2'-\vec x_j)\delta(M_1-M_i)\delta(M_2-M_j) \right>.
	\label{2point}
\end{align}
There are two contributions, 1-halo and 2-halo terms.

\paragraph{1-halo term}

Let us first consider the case of $i=j$ in the summation over halos. This means that $\vec x_1$ and $\vec x_2$ are inside the same halo. In this case the parenthesis in (\ref{2point}) becomes
\begin{align}
	\left< \sum_i \delta(\vec x_1'-\vec x_i) \delta(\vec x_2'-\vec x_i)\delta(M_1-M_i)\delta(M_2-M_i) \right>
	=\frac{dn_h}{dM_1} \delta(\vec x_1'-\vec x_2')\delta(M_1-M_2).
	\nonumber
\end{align}
By using this and by Fourier transforming the density profile $u_M$, we arrive at
\begin{align}
	\left<\delta_\rho(\vec x_1,z)\delta_\rho(\vec x_2,z)\right>_{\rm 1halo} =  \int \frac{d^3k}{(2\pi)^3}e^{i\vec k\cdot(\vec x_1-\vec x_2)} P_\rho^{\rm 1h}(k; z) .
	\label{deltarho_2p}
\end{align}
where the 1-halo power spectrum of the density fluctuation is given by
\begin{align}
	P_\rho^{\rm 1h}(k; z) = \frac{1}{\overline\rho^2_{m0}}\int dM\,M^2\frac{dn_h(z)}{dM} \left|u_M(k; z)\right|^2.
\end{align}

\paragraph{2-halo term}

Next let us consider the case of $i\neq j$, meaning that $\vec x_1$ and $\vec x_2$ belong to different halos.
In this case the parenthesis in (\ref{2point}) becomes the correlation of halo mass function:
\begin{align}
	&\left< \sum_i \delta(\vec x_1'-\vec x_i) \delta(M_1-M_i)
	\sum_{j} \delta(\vec x_2'-\vec x_j)\delta(M_2-M_j) \right> \nonumber 
	= \left<\frac{dn_h}{dM}(\vec x_1';M_1) \frac{dn_h}{dM}(\vec x_2';M_2) \right> \nonumber\\
	&=\frac{d\overline n_h(z,M_1)}{dM}\frac{d\overline n_h(z,M_2)}{dM} \left(1 + \left<\delta_h(\vec x_1',z; M_1)\delta_h(\vec x_2',z; M_2) \right> \right).
	\label{dddd_2h}
\end{align}
Here let us introduce the linear halo bias $b(M,z)$ such that~\cite{Cooray:2002dia}
\begin{align}
	\left<\delta_h(\vec x_1,z; M_1)\delta_h(\vec x_2,z; M_2) \right> = b(M_1,z) b(M_2,z)\left<\delta_\rho^{\rm (lin)}(\vec x_1,z)\delta_\rho^{\rm (lin)}(\vec x_2, z) \right>,
\end{align}
where $\delta_\rho^{\rm (lin)}(\vec x,z)$ denotes the linear density perturbation and
\begin{align}
	\delta_h(\vec x,z; M) = \frac{n_h(\vec x,z; M) - \overline n_h(z,M)}{\overline n_h(z,M)},
    ~~~~~~n_h(\vec x,z; M) \equiv M \frac{dn_h}{dM}(\vec x,z; M).
\end{align}
With the use of these quantities, we find\footnote{
	The term proportional to $1$ on the most right hand side of Eq.~(\ref{dddd_2h}) is canceled with the term $1$ on the left hand side of Eq.~(\ref{2point}).}
\begin{align}
	\left<\delta_\rho(\vec x_1,z)\delta_\rho(\vec x_2,z)\right>_{\rm 2halo} =  \int \frac{d^3k}{(2\pi)^3}e^{i\vec k\cdot(\vec x_1-\vec x_2)} P_\rho^{\rm 2h}(k;z).
\end{align}
where the 2-halo power spectrum is given by
\begin{align}
	P_\rho^{\rm 2h}(k;z) = \left[\frac{1}{\overline\rho_{m0}}\int dM M\frac{dn_h(z)}{dM} u_M(k; z) b(M,z)\right]^2 P_\rho^{\rm (lin)}(k; z).
\end{align}
The full nonlinear power spectrum is the sum of 1- and 2-halo terms:
\begin{align}
	P_\rho(k;z) = P_\rho^{\rm 1h}(k;z)  + P_\rho^{\rm 2h}(k;z).
\end{align}
In order to calculate these quantities we need expressions for the halo mass function $dn_h/dM$, the Fourier transformation of the halo density profile $u_M(k)$, linear halo bias $b(M)$, the linear matter power spectrum $P_\rho^{\rm (lin)}(k)$. They are summarized in Appendix~\ref{app:halo}.
The left panel of Fig.~\ref{fig:Pk} shows the dimensionless power spectrum, $\left(\frac{k^3}{2\pi^2}\right)\times \left(P^{\rm 1h}_\rho(k),P^{\rm 2h}_\rho(k),P^{\rm (lin)}_\rho(k),P_\rho(k)\right)$, all evaluated at $z=0$.

\subsection{Power spectrum for square matter density}
\label{app:4point}

With a discussion parallel to the previous subsection, the square matter density field $\rho_m^2(\vec x)$ is given by
\begin{align}
	\rho_m^2(\vec x,z) &= \sum_i \rho^2_h\left(\frac{\vec x-\vec x_i}{1+z}; z; M_i\right) = \sum_i (1+z)^6 M_i^2 u^2_{M_i}(\vec x-\vec x_i; z) \nonumber\\
	&=\int dM \int d^3x' \left[\sum_i \delta(\vec x'-\vec x_i) \delta(M-M_i)\right](1+z)^6 M^2 u^2_M(\vec x-\vec x'; z).
\end{align}
From this we immediately see
\begin{align}
	\left<\rho_m^2(\vec x,z)\right>=\Delta^2(z)\overline\rho_m^2(z) = (1+z)^{6}\int d^3x'\int dM\,M^2\frac{dn_h(z)}{dM}u_M^2(\vec x-\vec x';z).
\end{align}
It is the same as Eq.~(\ref{Delta2}). Its fluctuation is defined as $\delta_{\rho^2} = (\rho^2_m(\vec x,z)-\left<\rho_m^2(\vec x,z)\right>) / \left<\rho_m^2(\vec x,z)\right>$ (see Eq.~(\ref{delta_rho2})).
For evaluating the angular power spectrum from the dark matter annihilation, we also need to evaluate
\begin{align}
	&\left<\delta_{\rho^2}(\vec x_1,z)\delta_{\rho^2}(\vec x_2,z)\right> +1= \frac{1}{\Delta^4(z)\overline\rho_{m0}^4}\int dM_1\int dM_2\int d^3x_1'\int d^3x_2'
	\, M^2_1M^2_2 \nonumber \\
	&\times u^2_{M_1}(\vec x_1-\vec x_1';z)u^2_{M_2}(\vec x_2-\vec x_2';z)\left< \sum_{i,j}\delta(\vec x_1'-\vec x_i) \delta(\vec x_2'-\vec x_j)\delta(M_1-M_i)\delta(M_2-M_j) \right>.
\end{align}
Similar to the previous case, we have 1-halo and 2-halo terms. 

\paragraph{1-halo term}

Let us consider the case of $i=j$.
\begin{align}
	&\left<\delta_{\rho^2}(\vec x_1,z)\delta_{\rho^2}(\vec x_2,z)\right>_{\rm 1halo}
	= \int \frac{d^3k}{(2\pi)^3}e^{i\vec k\cdot(\vec x_1-\vec x_2)} P_{\rho^2}^{\rm 1h}(k; z), 
\end{align}
where
\begin{align}
	P_{\rho^2}^{\rm 1h}(k; z) = \frac{1}{\overline\rho^2_{m0} \Delta^4(z)}\int dM\,M^2\frac{dn_h(z)}{dM}\left| U_M(k;z) \right|^2.
\end{align}
Here we have defined $U_M(k; z)$ as
\begin{align}
	u_M^2(\vec x-\vec x'; z) \equiv \frac{\overline\rho_{m0}}{M} \int \frac{d^3k}{(2\pi)^3} e^{i\vec k\cdot(\vec x-\vec x')} U_M(k;z).
\end{align}
\begin{figure}\centering
\includegraphics[width=.48\textwidth]{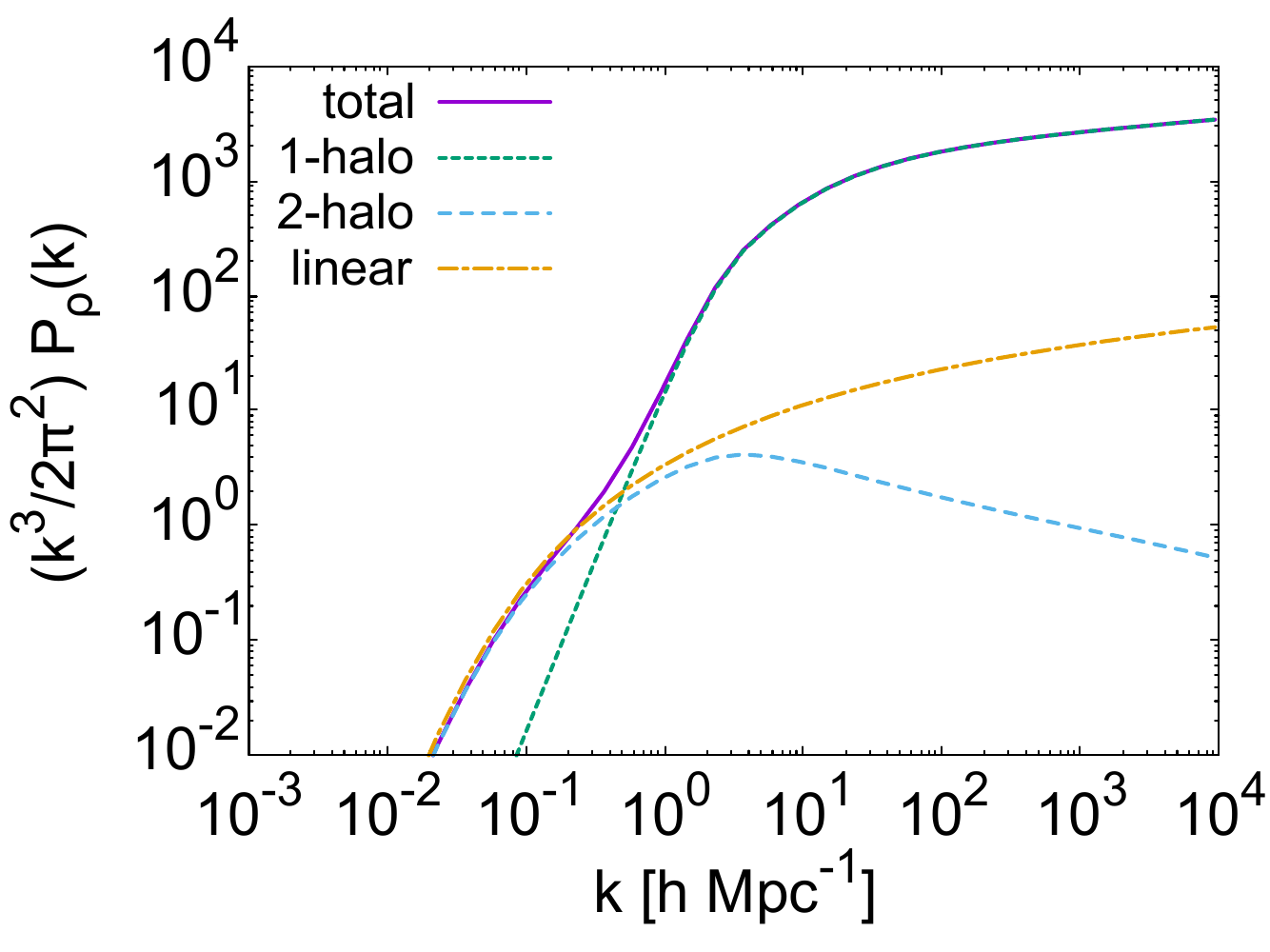}
\includegraphics[width=.48\textwidth]{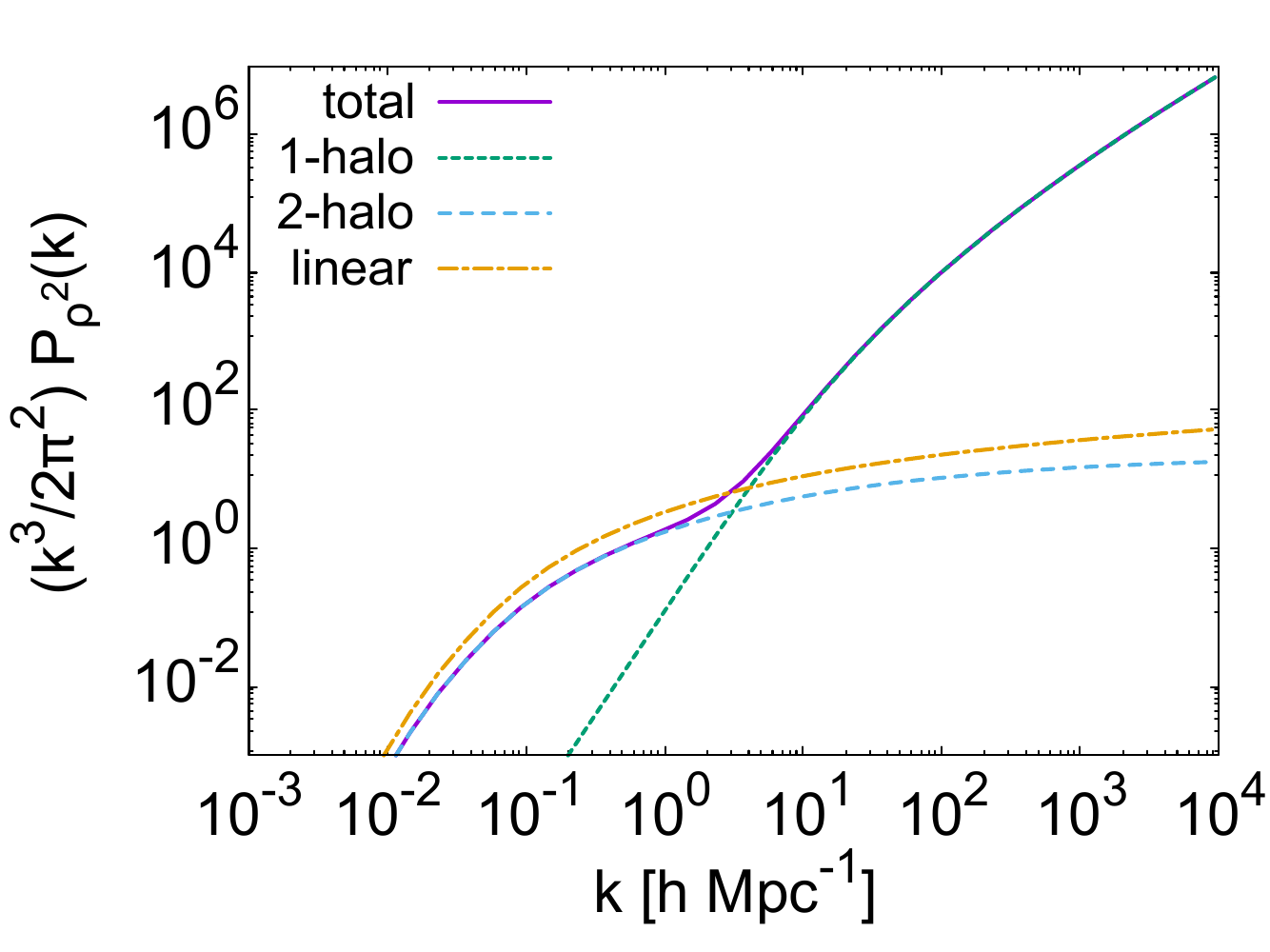}
\caption{The dimensionless nonlinear power spectrum of density fluctuation $\frac{k^3}{2\pi^2} P_{\rho}(k)$ (left) and the square density fluctuation $\frac{k^3}{2\pi^2} P_{\rho^2}(k)$ (right), both evaluated at $z=0$. In each figure, 1-halo, 2-halo and total spectra are separately shown. For reference the power spectrum of the linear density fluctuation is also shown.}
\label{fig:Pk}
\end{figure}

\paragraph{2-halo term}

The calculation of 2-halo term is also parallel. The result is 
\begin{align}
	P_{\rho^2}^{\rm 2h}(k;z) = \left[\frac{1}{\overline\rho_{m0} \Delta^2(z)}\int dM M\frac{dn_h(z)}{dM} U_M(k; z) b(M,z)\right]^2 P_\rho^{\rm (lin)}(k; z).
\end{align}
The full nonlinear power spectrum is the sum of 1- and 2-halo terms:
\begin{align}
	P_{\rho^2}(k;z) = P_{\rho^2}^{\rm 1h}(k;z)  + P_{\rho^2}^{\rm 2h}(k;z).
\end{align}
See Appendix~\ref{app:halo} for concrete expressions of the halo mass function $dn_h/dM$, Fourier transformation of the halo density profile $u_M(k)$, linear halo bias $b(M)$ and linear matter power spectrum $P_\rho^{\rm (lin)}(k)$.
The right panel of Fig.~\ref{fig:Pk} shows the dimensionless power spectrum, $\left(\frac{k^3}{2\pi^2}\right)\times \left(P^{\rm 1h}_{\rho^2}(k),P^{\rm 2h}_{\rho^2}(k),P^{\rm (lin)}_{\rho}(k),P_{\rho^2}(k)\right)$, all evaluated at $z=0$.

\section{Properties of dark matter halo} \label{app:halo}

\subsection{Halo mass function}

\begin{figure}\centering
\includegraphics[width=.5\textwidth]{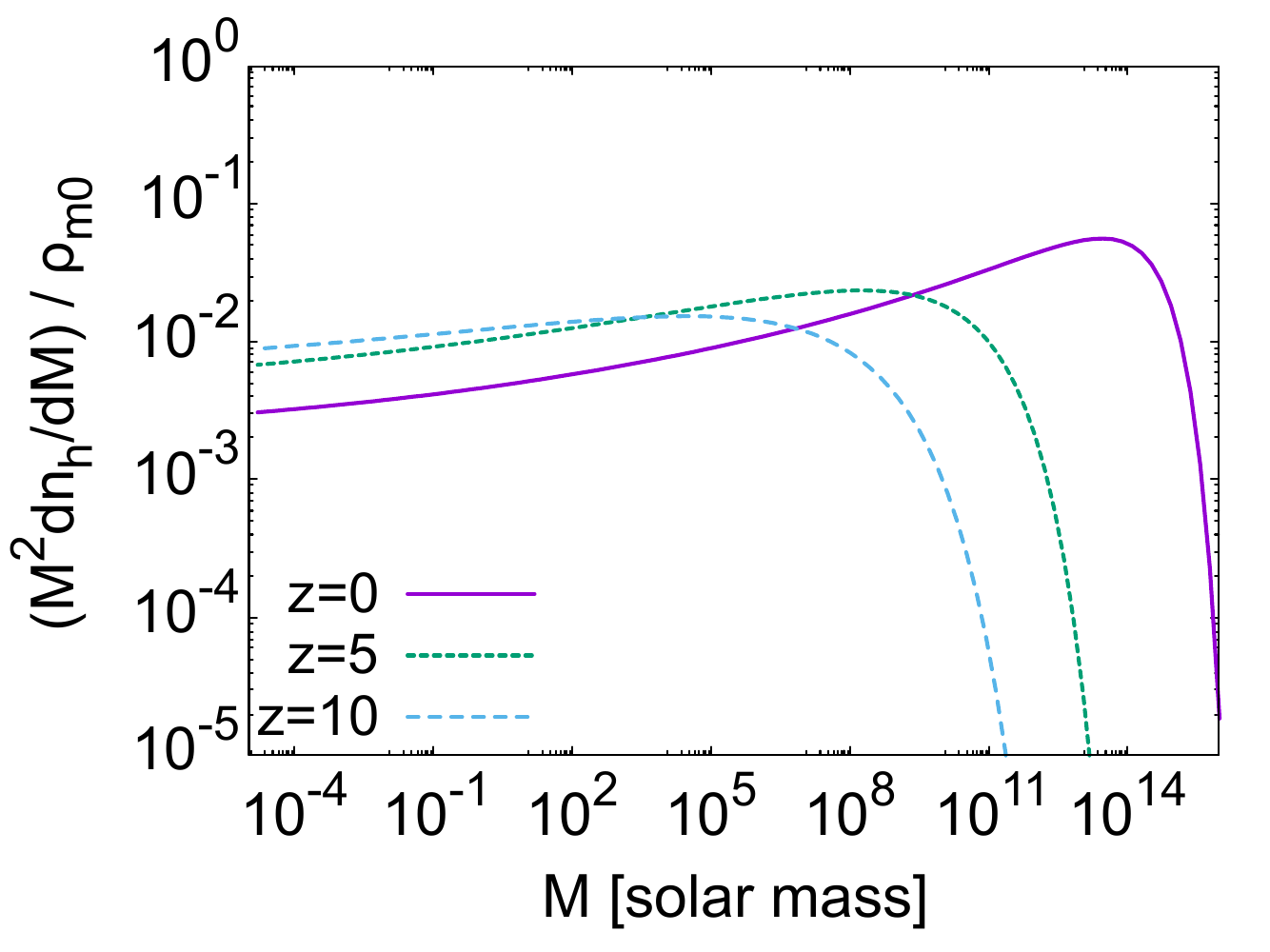}
\caption{The normalized halo mass function, $M^2\frac{dn_h}{dM}/\overline\rho_{m0}$ for $z=0,5,10$, are shown as a function of the halo mass $M$ in units of solar mass.}
\label{fig:dndM}
\end{figure}

\paragraph{Halo mass function} 
As for the dark matter halo mass function, we use the Sheth-Tormen formula~\cite{Sheth:1999mn} that takes into account the effect of ellipsoidal collapse:
\begin{align}
	\frac{dn_h(M;z)}{dM} = A\frac{\bar\rho_{m0}}{M^2}\left(\frac{\widetilde \nu}{2\pi}\right)^{\frac{1}{2}}\left(1+\frac{1}{\widetilde\nu^q}\right)e^{-\widetilde\nu/2}\frac{d\ln\nu}{d\ln M},
    \label{ST}
\end{align}
where the overall normalization constant is $A=0.322$ so that $\int dM\,M \frac{dn_h}{dM} = \overline\rho_{m0}$, $\nu=\delta_c^2(z)/\sigma^2(M)$, $\widetilde\nu = b\nu$ with $b=0.707$, $q=0.3$.\footnote{
    Mathematically one can integrate from $\nu=0$ to $\infty$ and obtain $A=0.322$ for correct normalization. In reality, the integral is sensitive to the lowest value of $\nu$ due to the mild $\nu$ dependence of the integrand and it is nontrivial whether $\nu$ can take such a small value or not in realistic cosmology.
}
The original Press-Schechter formula~\cite{Press:1973iz} based on the spherical collapse model is reproduced for $q\to 0,$ $b\to 1$ and $A\to 1/2$.\footnote{
    See Refs.~\cite{Bond:1990iw,Maggiore:2009rv} for derivation of the Press-Schechter formula in the excursion set formalism.
}
Below we present how to evaluate the variance $\sigma(M)$, critical overdensity $\delta_c(z)$ and the linear matter power spectrum $P_{\rho}^{\rm (lin)}(k;z)$.
The resulting halo mass function, in terms of $M^2\frac{dn_h}{dM}/\overline\rho_{m0}$, is shown in Fig.~\ref{fig:dndM} for $z=0,5,10$. This represents the fraction of halo with mass $M$ in the total matter density. 

\paragraph{Variance} 
The variance of the smoothed density fluctuation on the scale $R_M$ is defined by
\begin{align}
	\sigma^2(M) =\left<\widetilde\delta^2_\rho(\vec x, R_M) \right>,~~~
    \widetilde\delta_\rho(\vec x, R_M)\equiv \frac{3}{4\pi R_M^3}\int d^3 x'\delta_\rho(\vec x')\theta\left(1-\frac{|\vec x-\vec x'|}{R_M}\right),
\end{align}
where $R_M$ satisfies $\frac{4\pi}{3}R_M^3 \overline\rho_{m0}=M$. This is calculated as
\begin{align}
	\sigma^2(M) = \frac{1}{2\pi^2}\int dk\,k^2\left|\widetilde W(k R_M)\right|^2 P_\rho^{\rm (lin)}(k; z=0),
\end{align}
where $\widetilde W(x) = 3(\sin x-x\cos x)/x^3$ is Fourier transform of the top-hat window function.
Note that $\sigma(M)$ is usually defined at $z=0$, while $\delta_c(z)$ (explained below) is dependent on the redshift. One can equally regard $\sigma(M)$ as a redshift-dependent quantity while $\delta_c$ is constant.

\paragraph{Critical overdensity} 
The critical linear overdensity $\delta_c(z)$ is given by $\delta_c(z) = 1.67[D_1(z=0)/D_1(z)]$~\cite{Eke:1996ds}, where the linear growth factor $D_1(z)$ represents the growth of the density perturbation after the matter-radiation equality.
In the matter-dominated universe the density perturbation just grows like $\delta_\rho(z) \propto a(z)$ and hence $D_1(a)\propto a(z)$ but the cosmological constant changes the scaling. It is fitted by a formula~\cite{Eisenstein:1997jh}
\begin{align}
	D_1(z) = \frac{5\Omega_m(z)}{2(1+z)}\left[\left(1+\frac{\Omega_m(z)}{2}\right)\left(1+\frac{\Omega_\Lambda(z)}{70}\right)-\Omega_\Lambda(z) + \Omega^{4/7}_m(z) \right]^{-1},
\end{align}
where $\Omega_m(z) = \Omega_{m0}(1+z)^3/[\Omega_{\Lambda0}+\Omega_{m0}(1+z)^3]$ and $\Omega_{\Lambda}(z)=1-\Omega_m(z)$.
The overall normalization is fixed so that $D_1(z) \to a(z)$ for $a\to 0$.\footnote{
    Our definition ensures that the Bardeen potential $\Phi(k,t)$ satisfies $\Phi(k,t)=\left[D_1(z)/a(z)\right]\Phi_{\rm m}(k)$ where $\Phi_{\rm m}(k)$ represents $\Phi(k)$ evaluated after the matter-radiation equality for superhorizon modes.}

\paragraph{Linear power spectrum} 
The power spectrum of the linear density fluctuation is given by
\begin{align}
    P_\rho^{\rm (lin)}(k; z) = \frac{4}{25}\frac{2\pi^2}{k^3}\left(\frac{k^2}{\Omega_{m0} H_0^2}\right)^2\Delta_\zeta^2(k)T^2(k)D_1^2(z),
\end{align}
where $H_0$ is the present Hubble parameter, $\Delta^2_\zeta(k)$ is the dimensionless power spectrum curvature perturbation: $\Delta_\zeta^2(k) = (2.1\times 10^{-9})\times(k/k_*)^{n_s-1}$ with $k_*=0.05\,{\rm Mpc^{-1}}$ and $n_s = 0.96$~\cite{Planck:2018vyg}. The transfer function $T^2(k)$ represents the effect of non-growth of the density fluctuation for subhorizon scales in the radiation dominated era. It is normalized so that $T^2(k)\to 1$ for $k\ll k_{\rm eq}$ with $k_{\rm eq}$ being the comoving Hubble scale at the matter-radiation equality.  A fitting formula for $T^2(k)$ is found in Ref.~\cite{Eisenstein:1997jh}.

\paragraph{Linear halo bias}
For the estimation of 2-halo matter power spectrum, we need to know the expression for the linear bias factor $b(M;z)$ as discussed in the previous section. It is calculable once the halo mass function is given.
For the Sheth-Tormen halo mass function (\ref{ST}) it is given by~\cite{Sheth:1999mn,Cooray:2002dia}
\begin{align}
	b(M, z) = 1 + \frac{1}{\delta_c(z)}\left(\widetilde\nu-1+\frac{2q}{1+\widetilde\nu^q} \right).
\end{align}

\subsection{Halo density profile}
\label{app:density}

\paragraph{Density profile}
As for the density profile of the dark matter halo, we use the Navarro-Frenk-White (NFW) profile~\cite{Navarro:1996gj},
\begin{align}
	\rho_h(r; M; z) = \rho_s(M; z)\left[\frac{r}{r_s(M;z)}\left(1+\frac{r}{r_s(M;z)}\right)^2 \right]^{-1},
    \label{NFW}
\end{align}
where $r$ denotes the distance from the center of the halo.\footnote{One should not confuse it with the comoving distance $r(z)$ (\ref{rz}).} 
To determine $\rho_s(M;z)$ and $r_s(M;z)$, let us follow the procedure below.
The virial radius $R_{\rm vir}$ of the halo with its mass $M$ is defined as
\begin{align}
	\frac{4\pi}{3} R_{\rm vir}^3(M;z) \Delta_{\rm vir}(z) \overline \rho_m(z)= M,
\end{align}
where the expression for the virial overdensity $\Delta_{\rm vir}(z)$ in the presence of cosmological constant is presented in the next paragraph.
Let us define the concentration parameter $c_{\rm vir}$ as $c_{\rm vir}(M; z) \equiv R_{\rm vir}/r_{-2}$ where $r_{-2}$ is the distance at which the slope of the density profile becomes $-2$, i.e., $\left.\frac{d}{dr}\left(r^2\rho_h(r)\right)\right|_{r=r_{-2}}=0$. For the NFW profile, we have $r_{s}(M;z)=r_{-2} = R_{\rm vir}(M;z)/c_{\rm vir}(M;z)$. 
See the next paragraph for the estimation of $c_{\rm vir}(M;z)$.
Using these quantities, $\rho_s(M;z)$ is obtained by the condition
\begin{align}
	&M = \int_0^{R_{\rm vir}(M;z)} dr\,4\pi r^2  \rho_h(r; M; z)
    = 4\pi r_s^3(M;z)\rho_s(M;z) \mathcal F\left(c_{\rm vir}(M;z)\right),\nonumber\\
    & \mathcal F\left(c_{\rm vir}(M;z)\right) \equiv \log\left(1+c_{\rm vir}(M;z)\right)-\frac{c_{\rm vir}(M;z)}{1+c_{\rm vir}(M;z)}.
\end{align}
More explicitly, we have
\begin{align}
	\rho_s(M;z) = \frac{c^3_{\rm vir}(M;z)}{3 \mathcal F\left(c_{\rm vir}(M;z)\right)} \Delta_{\rm vir}(z) \overline\rho_m(z).
\end{align}
\begin{figure}\centering
\includegraphics[width=.48\textwidth]{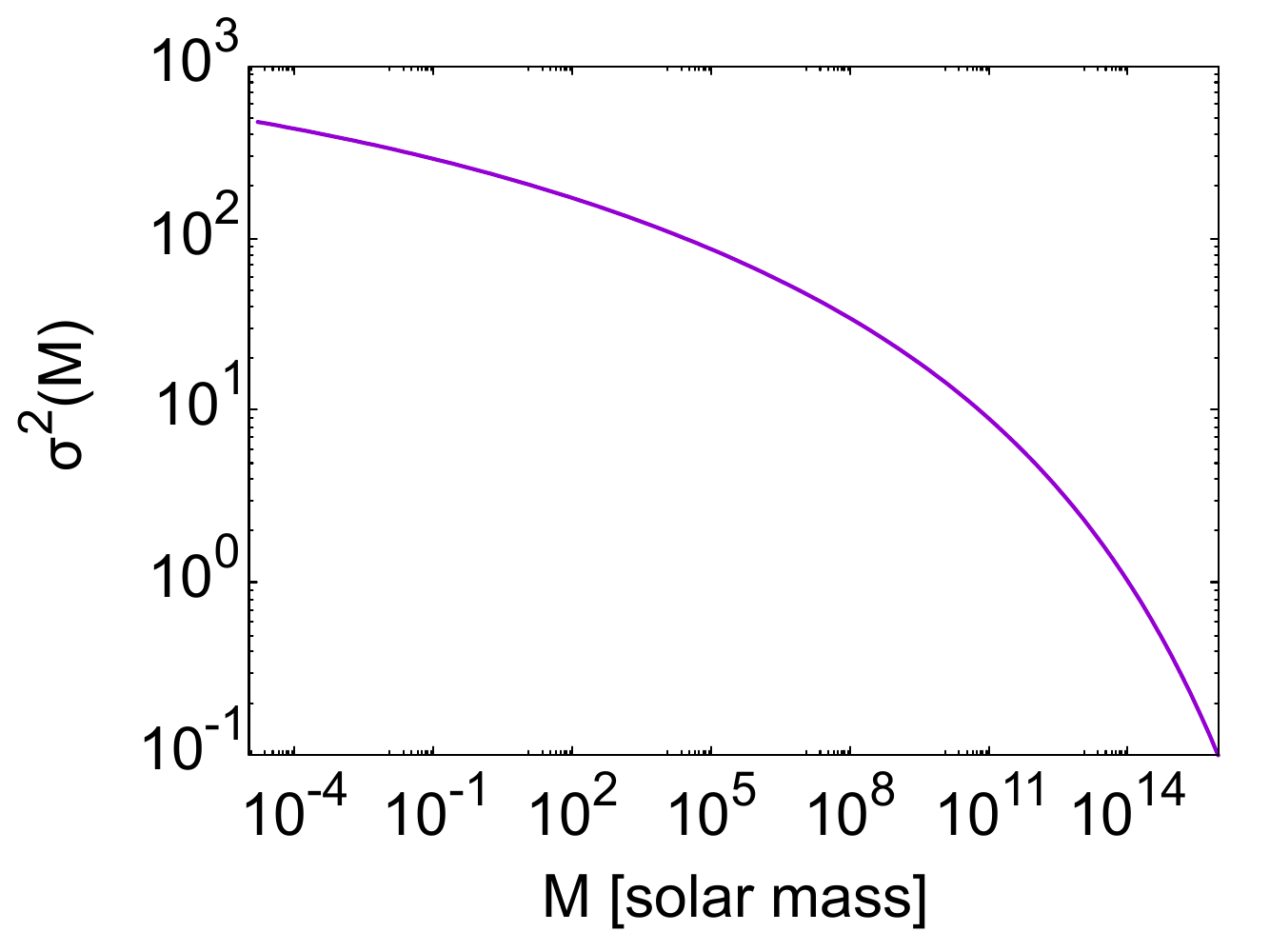}
\includegraphics[width=.48\textwidth]{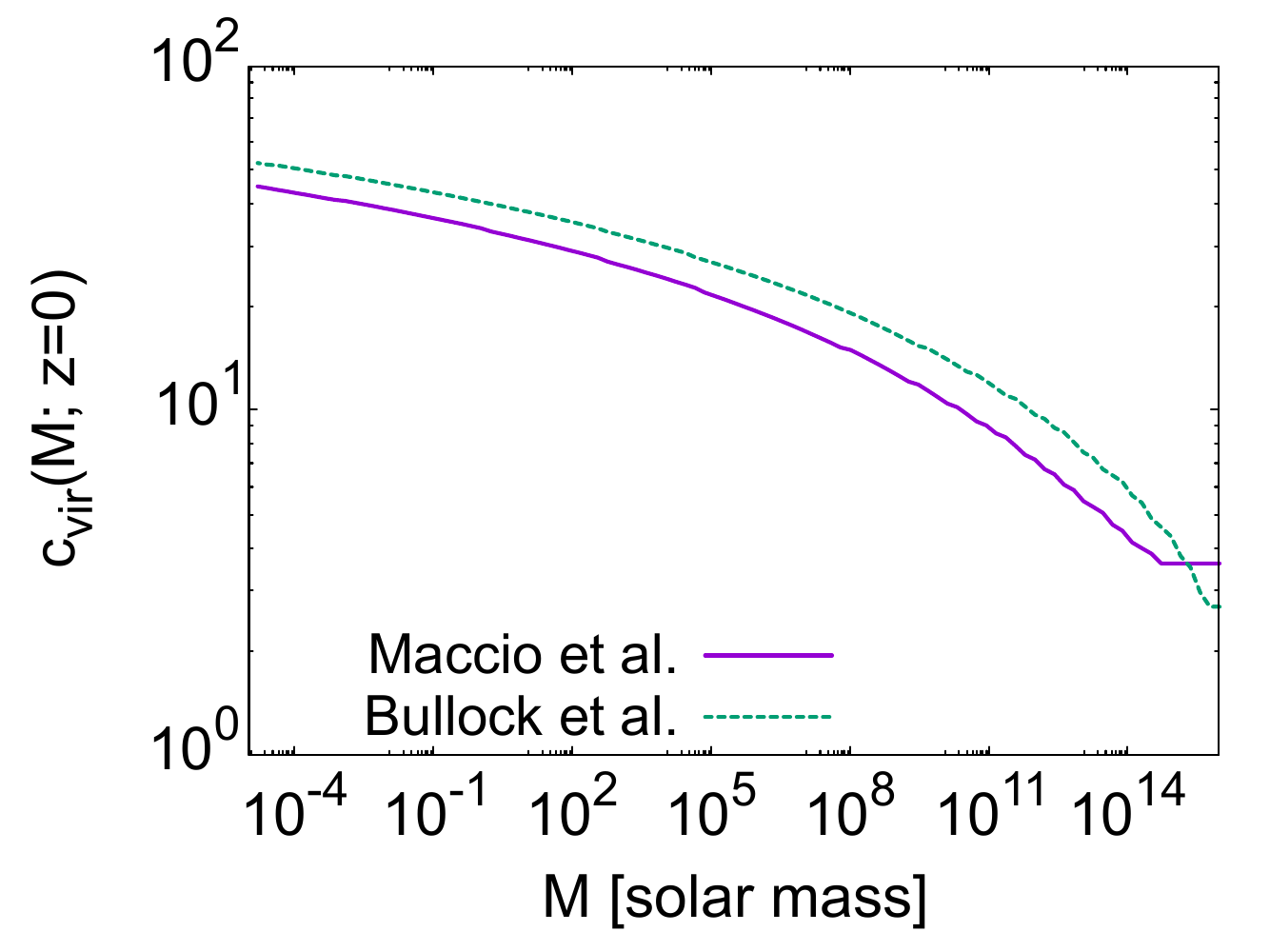}
\caption{(Left) The variance of the linear density perturbation $\sigma^2(M)$. (Right) The concentration parameter $c_{\rm vir}(M; z=0)$ for two different prescriptions described in the text. Both are plotted against the halo mass $M$ in units of solar mass.}
\label{fig:sigma}
\end{figure}

\paragraph{Virial overdensity and concentration parameter}
To complete the calculation, we need concrete expressions for $\Delta_{\rm vir}(z)$ and $c_{\rm vir}(z)$.
We consider two prescriptions for them. One is taken from Bullock et al.~\cite{Bullock:1999he}:
\begin{align}
	c_{\rm vir}(M;z) = K \frac{1+z_c(M)}{1+z},~~~~~~
    \Delta_{\rm vir}(z) = \frac{18\pi^2 -82\Omega_\Lambda(z) - 39 \Omega_\Lambda(z)^2}{\Omega_m(z)}.
    \label{cvir_Bullock}
\end{align}
where $K$ is a constant fitted by $N$-body simulation and $z_c(M)$ is the redshift at which the halo with mass $M$ was formed. 
It is calculated from the condition $\sigma(FM)=\delta_c(z_c)$ with another numerical constant $F$.
We adopt $K=2.7$ and $F=10^{-3}$, as written in Ref.~\cite{Maccio:2008pcd}.
The form of $\Delta_{\rm vir}(z)$ in Eq.~(\ref{cvir_Bullock}) has been derived in Ref.~\cite{Bryan:1997dn}.
Another prescription is taken from Maccio et al.~\cite{Maccio:2008pcd}:
\begin{align}
	c_{\rm vir}(M;z) = K\left[\frac{\Delta_{\rm vir}(z_c)\overline\rho_m(z_c)}{\Delta_{\rm vir}(z)\overline\rho_m(z)}\right]^{1/3},~~~~~~
    \Delta_{\rm vir}(z) = \frac{200}{\Omega_m(z)},
    \label{cvir_Maccio}
\end{align}
with $K=3.6$ and $F=0.01$. 
The right panel of Fig.~\ref{fig:sigma} shows $c_{\rm vir}(M;z=0)$ calculated with two prescriptions.
We basically use (\ref{cvir_Maccio}) throughout this paper unless otherwise stated.\footnote{
    We implicitly assumed $z_c>z$ in Eqs.~(\ref{cvir_Bullock}) or (\ref{cvir_Maccio}), but for large halos this assumption may break down. We simply take $c_{\rm vir}=K$ in such a case.}

\begin{figure}\centering
\includegraphics[width=.48\textwidth]{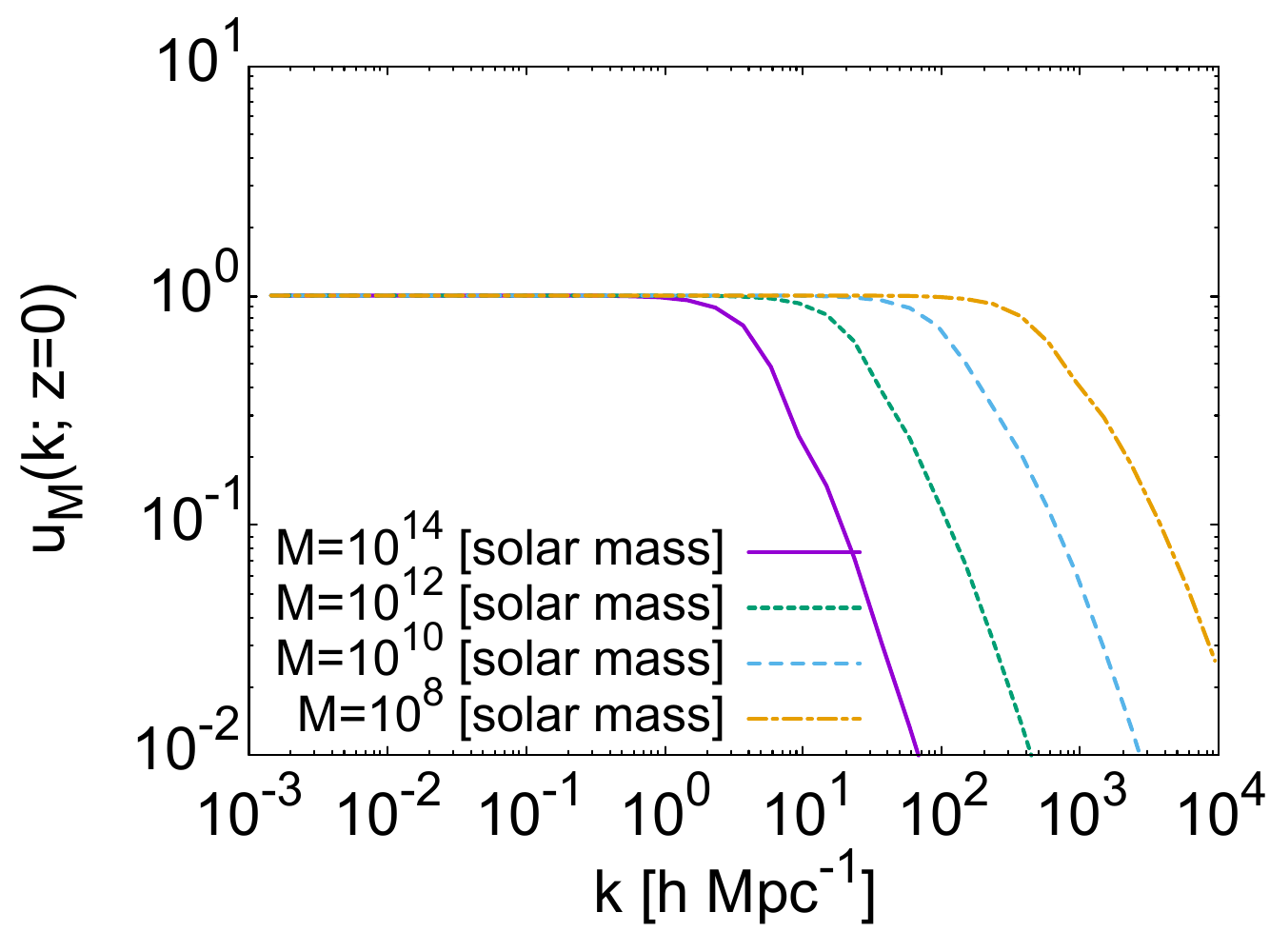}
\includegraphics[width=.48\textwidth]{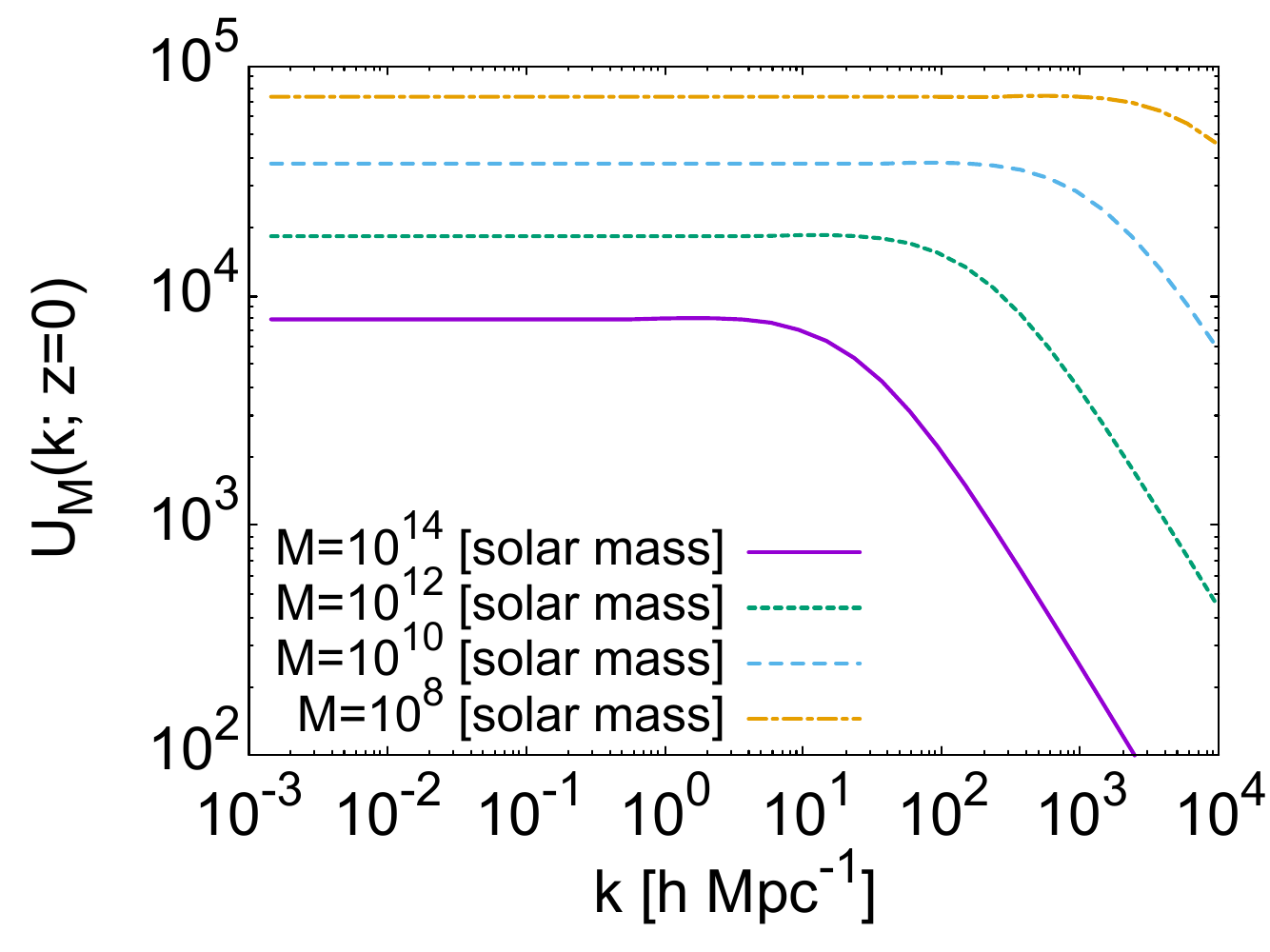}
\caption{The Fourier transformation of the halo density profile $u_M(k,z=0)$ (left) and of the square density profile $U_M(k,z=0)$ (right) as a function of the wave number $k$ in units of $h\,{\rm km/s/Mpc}$. Four lines correspond to $M/M_{\odot}=10^{14},10^{12},10^{10},10^8$.}
\label{fig:uMk}
\end{figure}

\paragraph{Fourier transformation of density profile}
The Fourier transformation of the density profile is given by
\begin{align}
	u_M(k;z)&=\int d^3x\,e^{-i\vec x\cdot\vec k} u_M(\vec x; z)
    =\int d^3x\,e^{-i\vec x\cdot\vec k}\frac{\rho_h(|\vec x|/(1+z); M)}{(1+z)^3 M} \nonumber\\
    & = \frac{4\pi r_s^3(M;z)\rho_s(M;z)}{M}\int_0^{\xi c_{\rm vir}}\frac{\sin y\,dy}{(\xi+y)^2}
    =\frac{1}{\mathcal F\left(c_{\rm vir}(M;z)\right)}\int_0^{\xi c_{\rm vir}}\frac{\sin y\,dy}{(\xi+y)^2} .
\end{align}
In the last line we have defined $y\equiv k|\vec x|$ and $\xi(M;z) \equiv (1+z)kr_s(M;z)$. 
Note that $u_M(k;z)\to 1$ in the $\xi\to 0$ limit.
The Fourier transformation of the square density profile is given by
\begin{align}
	U_M(k;z)&=\frac{M}{\overline\rho_{m0}}\int d^3x\,e^{-i\vec x\cdot\vec k} u^2_M(\vec x; z)
    =\frac{M}{\overline\rho_{m0}}\int d^3x\,e^{-i\vec x\cdot\vec k}\left[\frac{\rho_h(|\vec x|/(1+z); M)}{(1+z)^3 M}\right]^2 \nonumber\\
    & = \frac{4\pi r_s^3(M;z)\rho_s(M;z)}{M}\frac{\rho_s(M;z)}{\overline\rho_m(z)}\int_0^{\xi c_{\rm vir}}\frac{\xi^3\sin y\,dy}{y(\xi+y)^4}\nonumber\\
    &=\frac{c^3_{\rm vir}(M;z)\Delta_{\rm vir}(z)}{3\mathcal F^2\left(c_{\rm vir}(M;z)\right)}\int_0^{\xi c_{\rm vir}}\frac{\xi^3\sin y\,dy}{y(\xi+y)^4}.
\end{align}
Note that the prefactor in this expression is the same order as $\Delta^2(z)$ (\ref{Delta2}).
In Fig.~\ref{fig:uMk}, $u_M(k,z=0)$ and $U_M(k,z=0)$ are shown for $M/M_{\odot}=10^{14},10^{12},10^{10},10^8$.

\bibliographystyle{utphys}
\bibliography{ref}

\end{document}